\documentclass[]{aa}  

\usepackage{graphicx}
\graphicspath{ {images/} }
\usepackage{txfonts}
\usepackage{braket}
\usepackage{lipsum} 
\usepackage{tabularx}
\usepackage{ amssymb }
\usepackage{subcaption}
\usepackage{filecontents}
\newcommand{\rom}[1]{\uppercase\expandafter{\romannumeral #1\relax}}

\begin{document}

   \title{The EBLM project$^{\star}$}

   \subtitle{\rom{6}. The mass and radius of five low-mass stars in F+M binaries discovered by the WASP survey}

   \author{S. Gill
          \inst{1,2}
          \and
          P. F. L. Maxted
          \inst{1}
          \and
          J. A. Evans
          \inst{1}
          \and
          D. F. Evans
          \inst{1}
          \and
          J. Southworth
          \inst{1}
          \and
          B. Smalley
          \inst{1}
          \and
          B. L. Gary
          \inst{3}
          \and
          D.\ R.\ Anderson
          \inst{1}
          \and
          F.\ Bouchy
          \inst{4}          
          \and
          A.\ C.\ Cameron
          \inst{5}
          \and
          M.\ Dominik
          \inst{5}
          \and
          F. Faedi
          \inst{2}
          \and
          M.\ Gillon
          \inst{6}
          \and 
          Y.\  Gomez Maqueo Chew
          \inst{7}
          L.\ Hebb
          \inst{8}
          \and
          C.\ Hellier
          \inst{1}
          \and 
          U.\ G.\ J{\o}rgensen
          \inst{9}
          \and
          P.\ Longa-Pe\~na
          \inst{10}
          \and
          D.\ V.\ Martin
          \inst{11}
          \and
          J.\ McCormac 
          \inst{2}
          \and
          F.\ V.\ Pepe
          \inst{4}
          \and
          D.\ Pollaco
          \inst{2}
          \and
          D.\ Queloz
          \inst{4,12}
          \and
          D.\ S\'{e}gransan
          \inst{4}
          \and 
          C.\ Snodgrass
          \inst{13,14}
          \and 
          O.\ D.\ Turner
          \inst{4}
          \and
          A. H. M. Triaud
          \inst{15}
          \and
          S. Udry
          \inst{4}
          \and
          R.\ G.\ West
          \inst{2}
          }
          
\offprints{Samuel.Gill@warwick.ac.uk}

\institute{Astrophysics Group, Keele University,  Staffordshire, ST5\,5BG, UK
\and
Dept. of Physics, University of Warwick, Gibbet Hill Road, Coventry CV4 7AL, UK
\and
Hereford Arizona Observatory, Hereford, AZ 85615, USA
\and
Observatoire de Gen\'{e}ve, 51 chemin des maillettes, CH-1290 Sauverny, Switzerland
\and
Centre for Exoplanet Science, SUPA, School of Physics and Astronomy, University of St Andrews, St Andrews KY16 9SS, UK
\and
Astrobiology Research Unit, Université de Liège, 19C Allée du 6 Août, 4000 Liège, Belgium 
\and
Instituto de Astronom\'{i}a, Universidad Nacional Aut\'{o}noma de M\'{e}xico, Ciudad Universitaria, Ciudad de M\'{e}xico, M\'{e}xico
\and 
Department of Physics, Hobart and William Smith Colleges, Geneva, New York, 14456, USA
\and
Niels Bohr Institute \& Centre for Star and Planet Formation, University of Copenhagen {\O}ster Voldgade 5, 1350 - Copenhagen, Denmark         
\and 
Centro de Astronomía CITEVA, Universidad de Antofagasta, Av. Angamos 601, Antofagasta, Chile
\and 
Department of Astronomy and Astrophysics, University of Chicago, Chicago, IL 60637
\and
Cavendish Laboratory J J Thomson Avenue Cambridge, CB3 0HE, UK
\and 
School of Physical Sciences, Faculty of Science, Technology, Engineering and Mathematics, The Open University, Walton Hall, Milton Keynes, MK7 6AA, UK
\and
Institute for Astronomy, University of Edinburgh, Royal Observatory, Edinburgh EH9 3HJ, UK
\and
School of Physics and Astronomy, University of Birmingham, Birmingham, UK
        }

   \date{Received -; accepted-}




 
  \abstract
{
Some M-dwarfs around F-/G-type stars have been measured to be hotter and larger than predicted by stellar evolution models. Inconsistencies between observations and models need addressing with more mass, radius and luminosity measurements of low-mass stars to test and refine evolutionary models.
Our aim is to measure the masses, radii and ages of the stars in five low-mass eclipsing binary systems discovered by the WASP survey.
We use WASP photometry to establish eclipse-time ephemerides and to obtain initial estimates for the transit depth and width. Radial velocity measurements were simultaneously fitted with follow-up photometry to find the best-fitting orbital solution. This solution was combined with measurements of atmospheric parameters to interpolate evolutionary models and estimate the mass of the primary star, and the mass and radius of the M-dwarf companion. We assess how the best fitting orbital solution changes if an alternative limb-darkening law is used and quantify the systematic effects of unresolved companions. We also gauge how the best-fitting evolutionary model changes if different values are used for the mixing length parameter and helium enhancement.
We report the mass and radius of five M-dwarfs and find little evidence of inflation with respect to evolutionary models.  The primary stars in two systems are near the ``blue hook'' stage of their post sequence evolution, resulting in two possible solutions for mass and age. We find that choices in helium enhancement and mixing-length parameter can introduce an additional 3-5\,\% uncertainty in measured M-dwarf mass. Unresolved companions can introduce an additional 3-8\% uncertainty in the radius of an M-dwarf, while the choice of limb-darkening law can introduce up to an additional 2\% uncertainty.
The choices in orbital fitting and evolutionary models can introduce significant uncertainties in measurements of physical properties of such systems.}

  \titlerunning{EBLM project \rom{5}: five EBLMs discovered by the WASP survey}
    \authorrunning{Gill et al.}
    \maketitle


\section{Introduction}

Low-mass stars ($\leq$ $\,0.6\, \rm M_{\sun}$) have historically been challenging to study because of their intrinsic dimness and the low probability of finding them in eclipsing systems. Careful observations of double-lined eclipsing binaries (SB2s) can result in (almost) model-independent mass and radius estimates to a precision better than 3\% in some cases \citep{1991A&ARv...3...91A,2010A&ARv..18...67T}. Interferometry can be used to achieve similar results. Interferometrically determined visibility data can be fitted with a curve appropriate for a uniformly illuminated disk to determine an angular diameter. The resultant angular diameter must be corrected for the effects of limb-darkening to obtain the true angular diameter. This correction is subtle in the infrared but more prominent in the visible and depen on stellar atmospheric parameters \citep[see Figure 1 of ][]{2000MNRAS.318..387D}. Numerical values to correct a uniformly-illuminated angular diameter are presented in the form of coefficients for a given limb-darkening law \citep[e.g ][]{1974MNRAS.167..475H}. However, the use of finite bandwidths can introduce instrumental effects  \citep[known as \textit{bandwidth smearing} or \textit{chromatic aberration} in radio astronomy;  ][]{1989ASPC....6..247B} or inconsistent angular diameters across the band due to spectral lines \citep{1974MNRAS.167..475H}. 

\citet{2012ApJ...757..112B} acquired interferometric observations at the CHARA Array in the near-infrared $K'$ and $H$ bands \citep{2005ApJ...628..453T} for 21 nearby, single and bright red dwarfs. They measured radii with an uncertainty below 3\% and uncertainty in $T_{\rm eff}$ below 1\% and robustly demonstrate that models over-predict $T_{\rm eff}$ by $\sim3\%$, and under-predict radii by $\sim5\%$. \citet{2019MNRAS.484.2674R} used near-infrared long-baseline interferometry with \textit{PIONIER} at the Very Large Telescope Interferometer along with data from the second Gaia data release (Gaia DR2; \citealt{2018A&A...616A..10G}) to provide estimates for their linear radii, effective temperatures and luminosities. They show that Gaia underestimates M-dwarf temperatures by $\sim 8.2$\% and visually identify a discontinuity in the $T_{\rm eff}$-radius plane. They argue this discontinuity ($M_* \sim 0.23 M_\star$) arises during the transition from partly convective M-dwarfs to the fully convective domain despite residing in a less-massive regime than what is typically consider as the convective transition. \citet{2016AJ....152..141B} used white-light interferometric observations from the Hubble Space Telescope with radial velocity data from the McDonald Observatory to obtain astrometric solutions of M-dwarfs in binary systems. They achieved mass uncertainties as low as 0.4\% in some cases (median error of 1.8\%) but these objects are not transiting and so the radius of each component cannot be accurately inferred. Obtaining the mass for single interferometric stars often requires photometric calibrations for red-dwarfs \citep[e.g. ][]{1993AJ....106..773H,2000A&A...364..217D,2016AJ....152..141B,2019ApJ...871...63M}. The uncertainty attributed to mass-luminoisty relations has decreased significantly in recent years to around 2-3 \% in redder colours \citep{2019ApJ...871...63M} but makes it challenging to assess radius inflation for single stars observed with interferometry. 

A review of the literature by \citet{2018arXiv180505841C} found 90 stars less massive than 0.6\,M$_{\sun}$ of which mass and radius measurements are quoted to an accuracy of 10\% or better. These measurements show that some low-mass stars are hotter and larger than stellar evolutionary models predict \citep{Zhou2014, GomezMaqueoChewMoralesFaediEtAl2014}. The physics of low-mass stars is an interesting problem in its own right. M-dwarfs near the hydrogen burning limit are cool enough for their interiors to approach the electron Fermi temperature, resulting in an electron-degenerate gas where a classical Maxwellian description does not apply \citep{BaraffeChabrierBarmanEtAl2003}. Further complications arise from enhanced magnetic fields which may affect the evolution of M-dwarfs \citep{ChabrierGallardoBaraffe2007}. Large magnetic fields are thought to be 
induced by tidal interaction in close binaries, enhancing rotation and the dynamo mechanism. This inhibits convection in the core and may be responsible for inflating some stellar radii above those predicted by evolutionary models \citep[e.g. ][]{2006Ap&SS.304...89R, 2010A&ARv..18...67T, 2011ApJ...728...48K}. However, studies of single M-dwarfs with interferometry \citep{2012ApJ...757..112B} and those in double-lined eclipsing binaries \citep{2012ApJ...757...42F} are comparably inflated by around 3\% making it unclear whether tidal interactions can be blamed  \citep{Spada2013}. 

Using spectroscopy to measure the atmospheric composition of a low-mass star is challenging. Low surface temperatures permit molecules (such as TiO, VO, CaH and H$_{2}$O) to exist which manifest as a series of broad, and mostly blended lines in the optical part of the spectrum. This problem is eased slightly in the infrared, where there are some un-blended regions. This  allows equivalent-width measurements to be made of the Na\rom{1}, Ca\rom{1} and H$_{2}$O-K2 index in the K band \citep{Rojas-Ayala2012}. The most common technique for determining the temperature of M dwarfs is  comparing spectra to model atmospheres \citep{Gaidos2014}. The infrared flux method has also been extended from FGK stars to M dwarfs \citep{Casagrande2008}, however significant statistical deviations from interferometric temperatures  have been noted \citep{Mann2015}.

The Wide Angle Search for Planets  \citep[WASP; ][]{PollaccoSkillenCollierCameronEtAl2006}
is a survey for $0.8$--$ 2 \rm \, R_{Jup}$ objects transiting solar-like stars. Objects in this radius range can have masses which span three orders of magnitude, from Saturn-like planets to M-dwarfs. Consequently, WASP photometry has been used to identify hundreds of FGK stars with transiting M dwarf companions as a by-product of its successful exoplanet search. These systems are termed EBLMs (eclipsing binary, low-mass). We have invested considerable effort to characterise these systems, including hundreds of hours of telescope time to measure their spectroscopic orbits. A primary aim of the EBLM project is to improve our understanding of low-mass stars using accurate mass and radius measurements for transiting companions to FGK stars \citep{Triaud2013,GomezMaqueoChewMoralesFaediEtAl2014,Triaud2017, vonBoetticher2017, vonBoetticher2018}. So far, 13 systems have absolute parameters measured from follow-up photometry and a further 118 EBLMs presented by \citet{Triaud2017}  have secondary masses. Of these, two stars (WASP-30B and J1013$+$01) appears to be inflated, and a third (J0113+31) is measured to be $\sim$600\,K hotter than expected. This latter result comes from an analysis of the secondary eclipses in this system using infrared photometry. \citet{GomezMaqueoChewMoralesFaediEtAl2014} were unable to explain this result with any of the mechanisms discussed in their work, including tidal dissipation. A similar result was also seen for KIC1571511 \citep{Ofir2012} using high-precision optical photometry from the Kepler space telescope. In general, M-dwarfs measured within the EBLM project are consistent with stellar models within a few percent \citep{vonBoetticher2018}.

It is clear that we do not 
fully understand stars at the bottom of the main sequence and,  by implication, the planets orbiting them. With interest in low mass stars increasing 
as a result of recent exoplanet discoveries (e.g. TRAPPIST-1 and Proxima Centauri), more effort needs to be invested into understanding what makes some low-mass stars anomalous so that we can better understand the myriad of exoplanet systems that will be discovered with TESS \citep{Ricker2014}, MEarth \citep{2008AAS...212.4402C}, Speculoos \citep{GillonJehinDelrezEtAl2013} and eventually PLATO \citep{2016AN....337..961R}. In this paper we present high quality lightcurves and spectroscopic orbits of five EBLM systems and use these to measure the masses and radii of the stars in these systems to a precision of a few percent in some cases. Section \ref{observations} describes the origin and reduction of data used in this work, Section \ref{data_characterisation} details how system parameters we extracted and we present and discuss our results in Sections \ref{results}. We discus predictions from evolutionary models in Sections \ref{discuss:mass-radius} \& \ref{discuss:inflation} and possible sources of systematic uncertainty in Section \ref{discussion:systematics}.


\section{Observations and data reduction}\label{observations}

\begin{table*}
\caption{Summary of observations used to derive stellar atmospheric and orbital parameters. The square brackets indicate the filter corresponding to the preceding number of observations. We also present the SED measurements used in Sect. \ref{sed_fit}. }              
\label{Observeration_table}      
\centering                                    
\begin{tabular}{c l l  l l l l l l l l l l l l l }          

\hline
\hline
 & J2349$-$32 & J2308$-$46 & J0218$-$31 & J1847$+$39 & J1436$-$13 \\
\hline
J2000.0 \\
$\alpha$  & $23^{\rm h}49^{'}15.23^{"}$ & $23^{\rm h}08^{'}45.66^{"}$ & $02^{\rm h}18^{'}13.24^{"}$ & $18^{\rm h}47^{'}52.34^{"}$ & $14^{\rm h}36^{'}46.42^{"}$\\
$\delta$ & $-32^{\circ}46' 17.5^{"}$ & $-46^{\circ}06^{'}36.6^{"}$  & $-31^{\circ}05^{'}17.3^{"}$ & $+39^{\circ}58^{'}51^{"}$ & $-13^{\circ}32^{'}35.5^{"}$\\
Vmag & 11.53 & 11.36 & 9.96 & 11.73 & 12.48\\
\\
$Transit photometry$ \\
WASP  & 8144 & 14,369 & 7872 & 9639 & 53,259 \\
SAAO 1-m  & 345 [I] & 474 [R]  & - & -  & 136 [R]\\
CTIO & - & - & 78 [g'] &- &- \\
& & & 62 [z'] \\
& & & 71 [r'] \\
& & & 70 [z'] \\
HAO  & - & - & - & 605 [CBB] & - \\
& & & & 311 [g'] \\
& & & & 371 [z'] \\ \\

$Spectroscopy$ \\
CORALIE & 20 & 19 & 70 & -& 20\\
INT & - & - & - &10 &-\\
$\Delta t$ [yr] & 4 & 5 & 7.5 & 0.25 & 2\\ 
\\
$Gaia$ \\
$G$
& $11.448 \pm 0.001$
& $11.381 \pm 0.001$
& $ 9.775 \pm 0.001$
& $11.755 \pm 0.001$
& $12.334 \pm 0.001$\\

$G_{BP} - G_{RP}$
& $0.721 \pm 0.002$
& $0.728 \pm 0.002$
& $0.779 \pm 0.002$
& $0.818 \pm 0.002$
& $0.759 \pm 0.002$ \\

parallax [mas]
& $3.769 \pm 0.092$
& $2.187 \pm 0.113$
& $3.762 \pm 0.092$
& $3.583 \pm 0.086$
& $2.063 \pm 0.097$ \\\\

$photometry$ \\
APASS9 [B]  & 
$12.142 \pm 0.039$ &
$12.072 \pm 0.015$ &
$10.519 \pm 0.037$ &
$12.382 \pm 0.021$ &
$12.986 \pm 0.009$ \\

APASS9 [V]  & 
$11.541 \pm 0.010$ &
$11.517 \pm 0.045$ &
$9.903 \pm 0.026$ &
$11.913 \pm 0.022$ &
$12.480 \pm 0.014$ \\

APASS9 [g'] & 
$11.785 \pm 0.013$ &
$11.749 \pm 0.016$ &
$10.202 \pm 0.032$ &
$12.007 \pm 0.031$ &
$12.690 \pm 0.018$ \\

APASS9 [r'] & 
$11.438 \pm 0.033$ &
$11.382 \pm 0.014$ &
$9.779 \pm 0.029$ &
$11.704 \pm 0.006$ &
$12.354 \pm 0.021$ \\

APASS9 [i'] &
$11.317 \pm 0.013$ &
$11.286 \pm 0.006$ &
$9.632 \pm 0.079$ &
$11.548 \pm 0.006$ &
$12.231 \pm 0.064$ \\

TYCHO [B$_{\rm T}$] & 
$12.278 \pm 0.138$ &
$11.801 \pm 0.091$ &
$10.655 \pm 0.039$ &
$12.146 \pm 0.137$ &
- \\

TYCHO [V$_{\rm T}$] & 
$11.593 \pm 0.100$ &
$11.398 \pm 0.108$ &
$9.958 \pm 0.033$ &
$11.766 \pm 0.150$ &
- \\
2MASS [J] & 
$10.530 \pm 0.023$ &
$10.477 \pm 0.022$ &
$8.783 \pm 0.034$ &
$10.682 \pm 0.026$ &
$11.353 \pm 0.027$ \\

2MASS [H] & 
$10.249 \pm 0.022$ &
$10.270 \pm 0.024$ &
$8.555 \pm 0.031$ &
$10.362 \pm 0.032$ &
$11.040 \pm 0.021$ \\

2MASS [K$_{\rm S}$] & 
$10.184 \pm 0.019$ &
$10.166 \pm 0.020$ &
$8.493 \pm 0.025$ &
$10.306 \pm 0.021$ &
$10.987 \pm 0.019$ \\

DENIS [I$_{\rm C}$] & -  & - & - & - & $11.790 \pm 0.030$ \\
DENIS [J] & -  & - & - & - & $11.371 \pm 0.070$ \\
DENIS [K$_{\rm S}$] & -  & - & - & - & $10.912 \pm 0.070$ \\

NED [E(B-V)] & 
$0.010 \pm 0.034$ &
$0.007 \pm 0.034$ &
$0.024 \pm 0.030$ &
$0.088 \pm 0.030$ &
$0.072 \pm 0.034$ \\

\hline

\end{tabular}
\end{table*}

This section describes the data we have used to measure the physical properties of five EBLM systems (J2349$-$32, J2308$-$46, J0218$-$31, J1547$+$39, J1436$-$13) discovered by the WASP project (Sect. \ref{WASP}). The quality of WASP photometry is not good enough to measure masses and radii of the components to the desired precision of a few percent so we obtained more precise follow-up photometry to provide improved size estimates of both EBLM components. A summary of observations can be found in Table \ref{Observeration_table} and the dates and time of spectroscopic observations are detailed in Table \ref{appendix:spec_observations}.

The larger stars in these systems are far brighter than their M-dwarf companions and only the reflex motion of the primary star can be measured. These radial velocity measurements provide a constraint on the mass ratio 
and the analysis of the transit provides information on the relative sizes of the stars, but one additional independent constraint is needed to uniquely determine the scale of the binary system. The additional constraint we use is the mass of the primary star based on its density, effective temperature (T$_{\rm eff}$) and metallicity ([Fe/H]) estimated from stellar models. Values of T$_{\rm eff}$ and [Fe/H] come from analysis of the coadded spectra for each star (Sect. \ref{atmospheric_parameters}).

\subsection{Photometric colours used for SED fitting}\label{observations:sed}

Photometric colours for each target was extracted from the following catalogues: B$_{\rm T}$ and V$_{\rm T}$ magnitudes from the Tycho-2 catalogue \citep{2000A26A...355L..27H}; B, V, g$^{\prime}$, r$^{\prime}$ and i$^{\prime}$ magnitudes from data release 9 of the AAVSO Photometric All Sky Survey (APASS9; \citealt{2016yCat.2336....0H}; J, H and K$_{\rm s}$ magnitudes from the Two-Micron All-Sky Survey (2MASS; \citealt{2006AJ....131.1163S}; i$^{\prime}$, J and K magnitudes from the DEep Near-Infrared Southern Sky Survey (DENIS; \citealt{1997Msngr..87...27E}). The reddening maps by \citet{2011ApJ...737..103S} were used to estimate the total line-of-sight extinction in the direction of each target, ${\rm E}({\rm B}-{\rm V})_{\rm map}$. Values of ${\rm E}({\rm B}-{\rm V})_{\rm map}$ were calculated using the NASA/IPAC Extragalactic Database (NED) operated by the Jet Propulsion Laboratory, California Institute of Technology\footnote{https://ned.ipac.caltech.edu/help/extinction\_law\_calc.html}. Not all EBLMs have photometry in all catalogues; those that do are reported in Tables \ref{Observeration_table}.

\subsection{Gaia colours and Interferometry used to estimate distance and evolutionary status}

\begin{figure}
    \centering
    \includegraphics[width=0.5\textwidth]{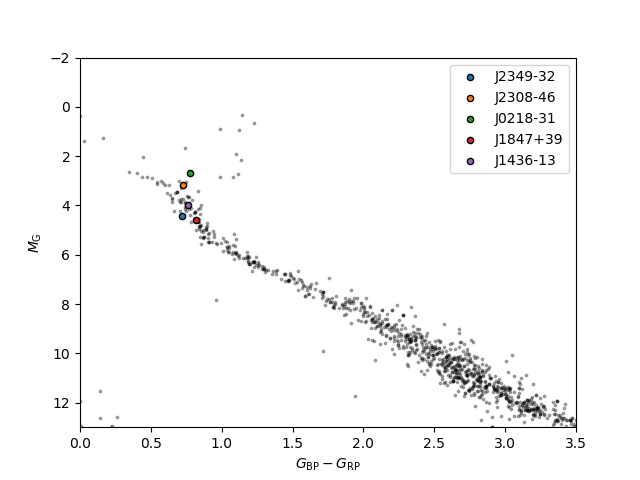}
    \caption{The $\rm M_{\rm G}$ VS $\rm G_{\rm BP}- \rm G_{\rm RP}$ plane for 160 randomly selected source fields (black) filtered using Eqns. 1 \& 2 from \protect\citet{2018A&A...616A..17A}. The EBLMs used in the work are also plotted and coloured appropriately. }
    \label{methods:fig:gaia_EBLMsmy_label}
\end{figure}


The second Gaia data release provides mean flux counts in three bands -- $G$, $G_{\rm BP}$ and $G_{\rm RP}$. The $G$-band has a wider wavelength coverage and is optimised to determine astrometric solutions. The mean magnitudes $G_{\rm BP}$ and $G_{\rm RP}$ provide a ``slice'' through the spectral energy distribution of stars and reveal how red or blue a star is. We obtained the mean $G$, $G_{\rm BP}$ and $G_{\rm RP}$ magnitudes along with parallax measurements for all nine EBLM systems from Gaia DR2 using the Gaia archive\footnote{https://gea.esac.esa.int/archive} (Table \ref{observations}). There is evidence of systematic offsets in parallax measurements from Gaia DR2 (e.g. \citealt{2018ApJ...862...61S}) which is likely correlated with on-sky positions ($\alpha$ \& $\delta$), $G$ and $G_{\rm BP}$ - $G_{\rm RP}$ \citep{2018A&A...616A...2L}. We added a systematic zero-point offset of -0.082\,mas to the parallax and added an additional 0.033\,mas in quadrature to the quoted parallax uncertainty \citep{2018ApJ...862...61S}. We plotted the position of all EBLMs in the $M_{\rm G}$-$G_{\rm BP}-G_{\rm RP}$ plane using data from 160 randomly selected source fields (Fig. \ref{methods:fig:gaia_EBLMsmy_label}) filtered using Eqns. 1 \& 2 from \citet{2018A&A...616A..17A}.

\subsection{WASP photometry for initial transit parameters and ephemerides}\label{WASP}

The WASP survey \citep{PollaccoSkillenCollierCameronEtAl2006} operates two survey instruments: one at the South African Astronomical Observatory (SAAO), South Africa, and another at the Observatorio del Roque de los Muchachos, La Palma.  Each instrument consists of an equatorial fork mount with eight cameras with 200-mm lenses and 2k$\times$2k CCD detectors. Each camera coveres approximately 64 square degrees per exposure. The data are processed by a detrending algorithm which was developed from the \textsc{SysRem} algorithm of \citet{2005MNRAS.356.1466T} and that is described by \citet{2007MNRAS.375..951C}. In July 2012, lenses on the southern installation (WASP-South) were changed to 85-mm with f/1.2 to search for brighter exoplanet hosts \citep{Smith2014}. Data from 85-mm lenses were not used in this study.

Photometry from the WASP cameras can suffer from a large amount of scatter due to clouds, instrumental artefacts, scattered light and other non-optimal observing conditions. We cleaned the data by removing points that were not detrended in the standard WASP reduction pipeline and removed points more than 0.5\,mag. from the median magnitude of each star. Additional cleaning of the light curve was done by comparing each night of data to a phase-folded light curve binned into 500 phase bins. Any measurement $3$-$ \sigma$ or more from the mean in each bin was excluded. The entire night of data was excluded if more than a quarter of the night's data was excluded this way or if there are fewer than 10 observations. The binned light curve is then inspected by eye to further exclude bad data points.

\subsection{SAAO 1-m follow-up transit photometry}

The SAAO hosts an equatorial-mounted 1-m telescope built by Grubb and Parsons that is equipped with an STE4 CCD camera with 1024\,$\times$\,1024 pixels. This camera was operated in $2 \times 2$ binning mode to reduce readout time. J2349$-$32 was observed on 18 October 2016 and J2308$-$46 on  12 October 2016 using $I$ (exposure time of $t_{\rm exp}$ = 50\,s) and $R$ ($t_{\rm exp}$ = 40\,s) Bessel filters. J1436$-$13 was observed on 23 April 2017 in the $R$ ($t_{\rm exp}$ = 40\,s) Bessel filter. Photometry was extracted using standard aperture photometry routines \citep{Southworth2009} and uncertainties were estimated from photon counting statistics. A by-eye approach was used to clean the light curve and select the best comparison star in the $5' \times 5'$ field. A slow variation in differential magnitude with time was observed corresponding to changes in the effective airmass. To correct for this, out-of-transit regions were manually selected and we used the IDL/AMOEBA\footnote{http://www.harrisgeospatial.com/docs/AMOEBA.html} routine to fit a polynomial which minimised the square of the magnitude residuals. The lightcurves were divided by the fitted polynomials to normalise to zero differential magnitude.

\subsection{HAO follow-up transit photometry}

\begin{figure}[htb]

  \centering
  \includegraphics[width=\textwidth/2]{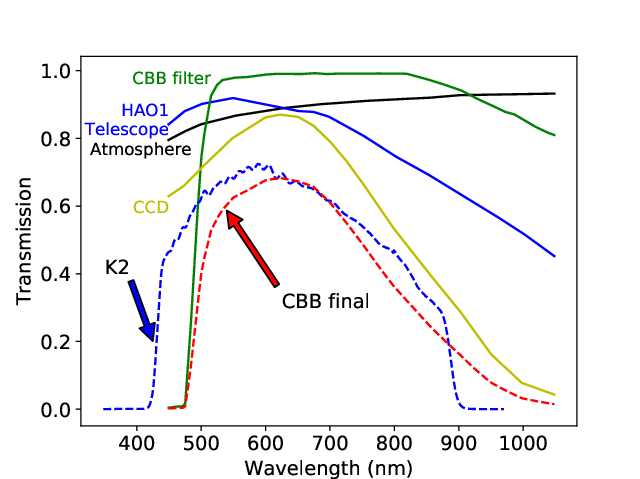}
  \caption{The response function of the HAO+CBB instrument. The atmospheric transmission is plotted in black, the transmission of the HAO telescope in blue-solid, the CBB filter in green and CCD response in yellow. The final response of HAO-1 with the CBB filter is plotted in red-dashed along with the Kepler transmission (blue-dashed). The atmospheric transmission line originated from equations for Rayleigh, aerosol and ozone extinction vs. wavelength for Palomar Observatory \protect\citep{1975ApJ...197..593H}. Coefficients were adjusted until they agreed with observations of extinction at HAO over a few dates (2018, priv. comm).}
  \label{HAO_CBB}
\end{figure}

\begin{figure}

  \centering
  \includegraphics[width=\textwidth/2]{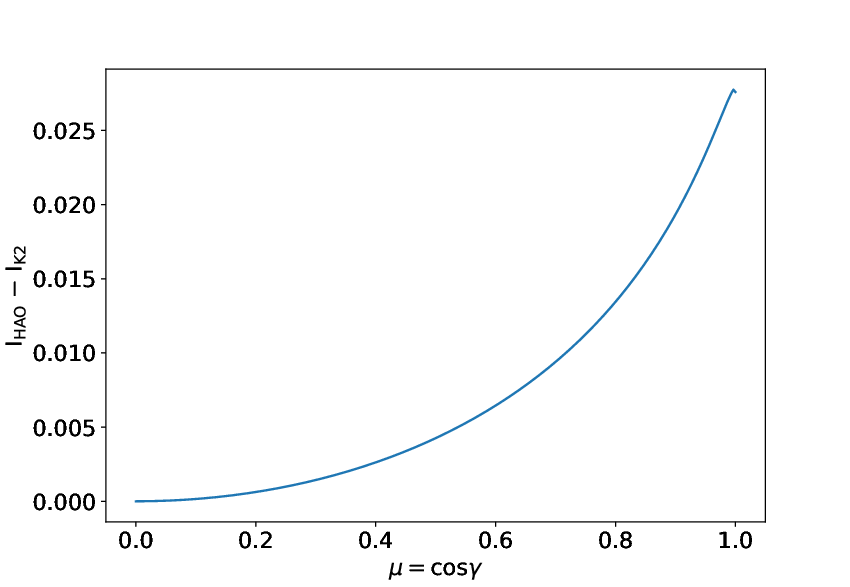}
  \caption{The difference in theoretical intensity acoss the stellar disk for the HAO+CBB filter and the Kepler/K2 Filter as a function of the angle between a line normal to the stellar surface and the line of sight of the observer ($\gamma$) for J1847$+$39. }
  \label{kepVShao}
\end{figure}

Optical photometry for J1847$+$39 was obtained at the Hereford Arizona Observatory, Arizona (HAO). Three separate transits were observed with a Meade 14-inch LX200GPS telescope. The first was obtained with the clear blue-blocking filter (CBB) on 9 October 2009 with $t_{\rm exp} = 100$\,s. The second was with a $g'$ filter on 18 May 2011 with $t_{\rm exp} = 60$\, s. The last was with a $z'$ filter on 15 June 2010 with $t_{\rm exp} = 60$\, s.  All observations were made in a binned mode of operation with a pixel scale of 1.5". The atmospheric seeing and telescope tracking produced a point-spread function with a full-width half-maximum of $\sim$4"(2.5 pixels). We chose apertures with radii of 4 pixels ($\sim 6$") to encapsulate the point-spread function of J1847$+$39 and comparison stars. Aperture photometry was extracted using standard photometry routines with systematic trends removed and outliers rejected.  Transmission information of the telescope throughput, atmosphere, filter and CCD\footnote{http://www.brucegary.net/HAO/} was used to calculate the final transmission of HAO with the CBB filter (see Fig.~\ref{HAO_CBB}). We used the four-parameter limb-darkening look-up table for the K2 passband instead of the CBB filter due to the similarity in final transmission since we do not have access to a four-parameter look-up table for the CBB filter. In Sect.~\ref{limb_darkening_section} we fit light curves using the two-parameter quadratic limb-darkening instead of the Claret law. The final response function in Fig.~\ref{HAO_CBB} was used  along with estimates of stellar atmospheric parameters from Sect. \ref{atmospheric_parameters} to calculate quadratic coefficients using \textsc{ldtk} \citep{Parviainen2015}. 

The discrepancy between the K2 and HAO+CBB pass-band differ in the blue where the limb-darkening is most significant. The validity of this assumption hinges on the fact that 1) the intensity across the stellar disc is similar in both the K2 and HAO+CBB bandpass and 2) the difference in limb-darkening coefficients for each band in negligable. The first assumption was tested using \textsc{ldtk} to synthesise intensity profiles for J1847$+$39 across the stellar disk for each pass-band and calculate the discrepancy as a function of $\gamma$ (the angle between a line normal to the stellar surface and the line of sight of the observer; Fig. \ref{kepVShao}). The K2 pass-band emits 2.5\,\% less flux than what would be observed with HAO+CBB towards the limb. The second assumption was tested by calculating quadratic limb-darkening coefficients for the K2 pass-band to be $u_1 = 0.496 \pm 0.050$, $u_2 = 0.157 \pm 0.050$ and for the HAO+CBB pass-band to be $u_1 = 0.468 \pm 0.050$, $u_2 = 0.148 \pm 0.051$. These are comparable within 1-$\sigma$ and so adopting the K2 pass-band for J1847$+$39 will have a negligible effect on the transit shape.

\subsection{CTIO follow-up transit photometry}

J0218-31 was observed on 14 November 2010 with the CTIO-0.9-m telescope and Tek2K CCD camera. The detector consists of a 2K$\times$2K array of $15\mu m$ pixels placed at Cassegrain focus giving a $0.4^{\prime\prime}$/pixel plate scale. Thus the entire array projects to a $13.7^{\prime}$ FOV. The observed signal is fed into four amplifiers causing the raw images to have a quadrant effect with the readnoise between 3.9-4.5~$e^{-}$ and gain of 2.5-2.8~$e^{-}$/ADU, depending on the amplifier. The detector has a readout time of 39 seconds and a 60,000 count well depth before non-linearity sets in when using $1\times 1$ binning mode. J0218-31 and the surrounding field were monitored throughout the night using the Sloan $griz$ filter set alternating and continuously between all four filters. Exposure times were chosen to maximise the flux in the target star and nearby reference stars while keeping the peak pixel value in J0218$-$31 below $60,000$ counts (well depth).  The telescope was defocused to allow for longer exposure times to build up signal in the fainter reference stars without saturating J0218-31. We adopted an exposure time of 10~seconds for the $g^{\prime}$, $r^{\prime}$, and $i^{\prime}$--band observations and longer exposures of 15~seconds in the $z^{\prime}$ filter where the detector is less sensitive. An overall light curve cadence of approximately 3.3 minutes was achieved in each filter accounting for the exposure times,  the read out time, and filter changes. The light curves were created from approximately 75 images taken in each filter during the single observing night.

A set of 11 bias calibration frames and 11 dome flat fields in all four filters were  obtained at the beginning of the observing night. The images were processed in a standard way using routines written by L.\ Hebb in the IDL programming language. Each of the four amplifiers was processed independently. All object and calibration frames were first overscan corrected (by subtracting a line-by-line median overscan value), bias subtracted and then trimmed.  Stacked bias images were created by averaging all bias frames observed each night and subtracted from all science  and flat-field frames.  All dome flats were averaged into a single dome flat in each filter and then applied to the trimmed and bias-corrected science images. 

Source detection and aperture photometry were performed on all processed science images using the Cambridge Astronomical Survey Unit catalogue extraction software \citep{2001NewAR..45..105I}. The  software  has  been  compared  with  SExtractor \citep{1996A26AS..117..393B} and found to be very similar in the completeness, astrometry and photometry tests.\footnote{\url{https://www.ast.cam.ac.uk/ioa/research/vdfs/docs/reports/simul/index.html}} This photometry software was applied to all processed images of J0218$-$31.  Adopting conservative parameters to define the detection threshold, the target star and dozens of fainter stars in the field were detected in each image.  Aperture photometry was performed on all detected stars using a 5 pixel radius circular aperture, which was selected to match the typical seeing. Five bright, non-variable reference stars were selected from the many detected stars and used to perform differential photometry on the target star.  In each image, the flux from all reference stars was summed into a single \textit{super} comparison star that was divided by the aperture flux from J0218$-$31 and converted to a differential magnitude.

\subsection{CORALIE spectra used for radial velocities and atmospheric parameters}

CORALIE is a fiber-fed \'{e}chelle spectrograph installed on the 1.2-m Leonard Euler telescope at the ESO La Silla Observatory and has a resolving power R = 50,000\,--\,60,000 \citep{2001A&A...379..279Q,Wilson2008}. The spectra used in this study were all obtained with an exposure times between $t_{\rm exp} = 600$-900\,s. Observations of J0218$-$31 include spectra obtained through the primary eclipse that show the Rossiter-McLaughlin effect. The spectra for each star were processed with the CORALIE standard reduction pipeline \citep{26AS..119..373B}. Radial velocity measurements were obtained using standard cross-correlation techniques (using numerical masks) and checked for obvious outliers \citep{Triaud2017}. Each spectrum was corrected into the laboratory reference frame and co-added onto a common wavelength range. Maximum and median filters were applied to identify continuum regions which were fitted with spline functions (one every nm) to normalise the spectra (a standard function within \textsc{ispec} v20161118; \citealt{Blanco-Cuaresma2017}).

\subsection{INT spectra used for radial velocities and atmospheric parameters}

Spectra for J1847$+$39 were obtained using the intermediate dispersion spectrograph (IDS) mounted on the 2.5-m Isaac Newton telescope (INT) at the Roque de Los Muchachos Observatory. The 235-mm camera and EEV10 CCD detector was used with the H1800V grating to obtain spectra in a small region around the H$\alpha$ line with R$\approx$10,000\footnote{Calculated from http://www.ing.iac.es/}. A total of 10 spectra were obtained for J1847$+$39 with an exposure time $t_{\rm exp} = 600$-$900$\, s. Radial velocity measurements were extracted using cross-correlation routines provided within \textsc{ispec}. A synthetic F0 spectrum was used as a template with a mask applied to the core of the H$\alpha$ line. A Gaussian function was fitted to the peak in each cross-correlation function to obtain the radial velocity measurement (the peak of the Gaussian function), and uncertainty (standard deviation of the Gaussian function). Each spectrum was corrected into a laboratory reference frame and co-added onto a common wavelength range. The relatively small wavelength range does not permit the use of maximum and median filters to normalise the spectra. Suitable continuum regions were identified by-eye to normalise the spectrum using a second-order polynomial fit by least-squares.  

\subsection{Lucky imaging used to identify nearby companions}

The lucky-imaging technique (e.g. \citealt{2006A&A...446..739L}) was used to obtain high-resolution images of J2308$-$46, J2349$-$32, J0055-00, J1652-19 and J2217-04 in July 2017, in order to search for stars contributing contaminating light, as well as potential bound companions to the eclipsing binaries. The observations were conducted using the Two Colour Instrument (TCI) on the Danish 1.54-m Telescope at La Silla Observatory. The TCI consists of two Electron Multiplying CCDs capable of imaging simultaneously in two passbands at a frame rate of $10$\,Hz, with a $40"\times40$" field of view. The `red' arm has a passband similar to a combined $i+z$ filter or the Cousins $I$ filter, whilst the `visible' arm has a mean wavelength close to that of the Johnson $V$ filter. A detailed description of the instrument  can be found in \citet{2015A26A...574A..54S} and  the lucky imaging reduction pipeline is described by \citet{2012A&A...542A..23H}.

The observations and data reduction were carried out using the method outlined in \citet{2018A26A...610A..20E}, and is briefly described here. Both targets were observed for 170\,s. The raw data were reduced automatically by the instrument pipeline, which performs bias and flat frame corrections, removes cosmic rays, and determines the quality of each frame, with the end product being ten sets of stacked frames, ordered by quality. The data were run through a custom star-detection algorithm that is described in \citet{2018A26A...610A..20E}, which is designed to detect close companion stars that may not be fully resolved.


\section{Methods}\label{data_characterisation}


This section describes the methods we have used to analyse our data in order to measure the masses and radii of each EBLM system. Our method shares some similarities to \citet{vonBoetticher2018}; we use the same spectroscopic analysis routine from \citet{Gill2018} which was confirmed with SED fitting, the same light curve model and Bayesian sampling routines for the orbital solution (although some fitted parameters are different). We used a modified sampler from \citet{vonBoetticher2018} to measure the masses and radii of each EBLM system from common stellar models. Our approach to quantifying inflation is also different to \citet{vonBoetticher2018}, both in terms of the models we use and approach. For completeness, we describe our methods in the following sections.


\subsection{SED fitting}\label{sed_fit}

Empirical colour--effective temperature relations were used used to estimate the effective temperature of the primary star in each system. These were used to complement our spectroscopic analysis and to provide a measurement of reddening. They were not used to interpolate stellar models or inform limb-darkening coefficients. We also assume that the flux contribution from the M-dwarf companion is negligible compared to the primary star (see Sect. \ref{M_dwarf_luminoisity}).

 Our model for the observed photometry has the following parameters --
g$^{\prime}_{0}$: the apparent g$^{\prime}$-band magnitude corrected for extinction; $T_{\rm eff}$, the effective temperature; E$({\rm B}-{\rm V})$, the reddening to the system; and $\sigma_{\rm ext}$, 
the additional systematic error added in quadrature to each measurement to
account for systematic errors. For each trial combination of these parameters the empirical colour -- effective temperature relations of
\cite{Boyajian2013} were used to predict the apparent magnitudes of the  star
in each of the observed bands. The transformation between the
Johnson and 2MASS photometric systems is the same as Boyajian et al. 2013. The  Cousins I$_{\rm C}$ band was used as an approximation to the DENIS Gunn i$^{\prime}$ band and the 2MASS K$_{\rm s}$  as an approximation to the DENIS K band \citep[see Fig.
4 of][]{Bessell2005}. Table 3 of
\citet{Bessell2000} was interpolated to transform the Johnson B, V magnitudes to
Tycho-2 B$_{\rm T}$ and V$_{\rm T}$ magnitudes. This assumed that the extinction
in the V band is $3.1\times {\rm E}({\rm B}-{\rm V})$. Extinction in the SDSS
and 2MASS bands is calculated using A$_{\rm r} = 2.770\times {\rm E}({\rm
B}-{\rm V})$ from \cite{Fiorucci2003} and extinction coefficients
relative to the r$^{\prime}$ band taken from \citet{Davenport2014}.

The reddening maps by \citet{Schlafly2011} were used to estimate the total
line-of-sight extinction in the direction of each target, ${\rm E}({\rm B}-{\rm V})_{\rm map}$.
This value is used to impose the following (unnormalized) prior on $\Delta =
{\rm E}({\rm B}-{\rm V}) - {\rm E}({\rm B}-{\rm V})_{\rm map}$: \[ P(\Delta) =
\left\{ \begin{array}{ll} 1 & \Delta \le 0 \\ \exp(-0.5(\Delta/0.034)^2) &
\Delta > 0 \\ \end{array} \right. \] The constant 0.034 is taken from
\citet{Maxted2014} and is based on a comparison of ${\rm E}({\rm
B}-{\rm V})_{\rm map}$ to ${\rm E}({\rm B}-{\rm V})$ determined using Str\"{o}mgren
photometry for 150 A-type stars. We used {\sc emcee} \citep{Foreman-Mackey2013} to sample the posterior probability distribution (PPD) for our model parameters. {\sc{emcee}} uses affine-invariant ensemble sampling \citep[parallel stretch move algorithm; ][]{Goodman2010} to split Markov chains into sub-groups and update the position of a chain using the positions of chains in the other subgroups. The algorithms affine-invariance can cope with skewed probability distributions and generally has shorter autocorrelation times than a classic Metropolis-Hastings algorithm. The empirical colour--temperature relations we have used are valid over the
approximate range T$_{\rm eff} = 3450$\,K to 8600\,K. 
Between these limits uniform priors were used
on the values of T$_{\rm eff}$. We also use uniform priors for
g$^{\prime}_{0}$. We evolved 10,000 steps from 100 walkers as a burn-in. A further 10,000 steps were drawn and the step with the highest likelihood value is selected,  with uncertainties equal to the standard deviation of each parameter in the second chain. An example posterior probability distribution (PPD) for J2308$-$46 is shown in Fig. \ref{WASP2308-46_fitmag}; the  PPDs for the other targets are shown in the appendix. The residuals of each fit to all of the EBLMs (observed magnitudes - calculated magnitudes) are shown in Fig. \ref{SED_BEST_FIT}. 

\begin{figure}[htb]
  \centering
  \includegraphics[width=\textwidth/2]{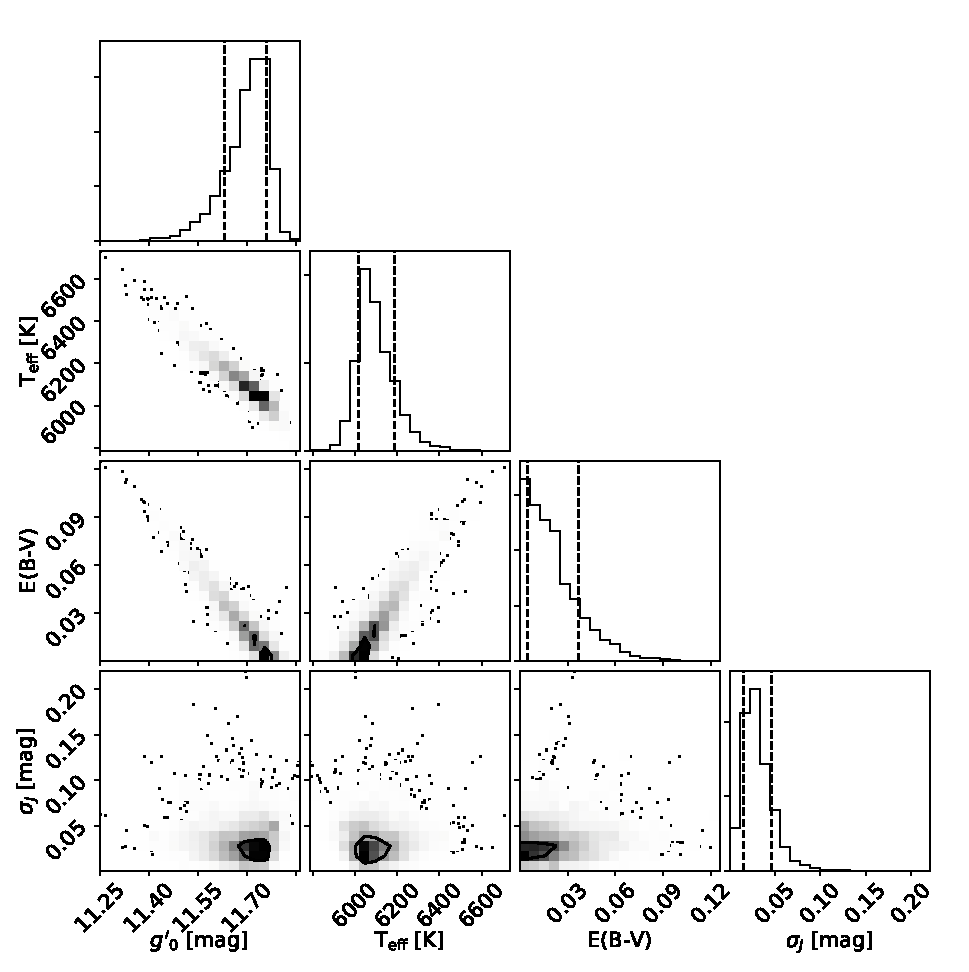}
  \caption{The posterior probability distribution of EBLM J2349$-$32 from photometric fitting. The 1-$\sigma$ contour are shown.}
  \label{WASP2308-46_fitmag}
\end{figure}

\begin{figure}[htb]
  \centering
  \includegraphics[width=\textwidth/2]{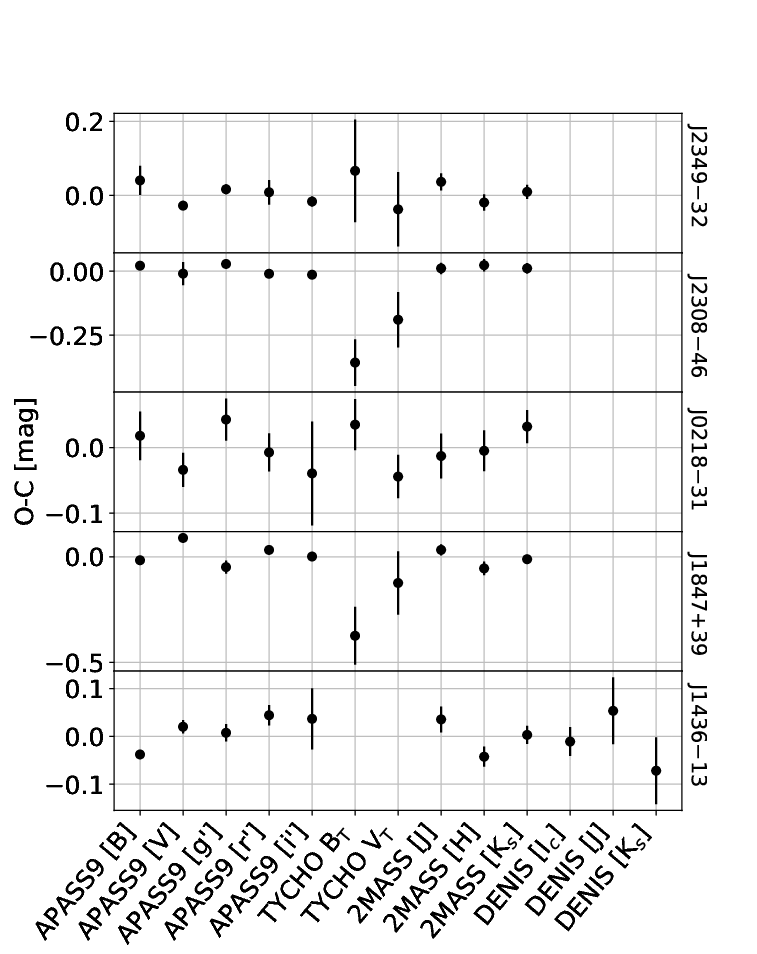}
  \caption{Difference between observed and calculated magnitudes for each EBLM. }
  \label{SED_BEST_FIT}
\end{figure}

\subsection{Spectroscopic analysis}\label{atmospheric_parameters}

In this section we describe how we measured $T_{\rm eff}$, [Fe/H], $V\sin i$ and $\log g$ from the CORALIE spectra or INT spectra.

\subsubsection{CORALIE - wavelet analysis}

We re-sample between 450-650\,nm with $2^{17}$ values and co-add the spectra. We calculate the wavelet coefficients $W_{i=4-14, k}$ (see Fig. 2 of \citet{Gill2018} for visual justification of our choice of wavelet coefficients) and fit the same coefficients with model spectra in a Bayesian framework. We initiated 100 walkers and generated 100,000 draws as a burn-in phase. We generated a further 100,000 draws to sample the PPD for $T_{\rm eff}$, [Fe/H], $V\sin\,i$ and $\log g$. \citet{Gill2018} note an [Fe/H] offset of $-0.18$\,dex which we correct for by adding 0.18 dex to the PPD for [Fe/H]. They also note a  significant trend in $\log g$ with $T_{\rm eff}$ which we also correct for using their Eqn. 9. The wavelet method for CORALIE spectra can determine $T_{\rm eff}$ to a precision of $85$\,K, [Fe/H] to a precision of 0.06\,dex and $V \sin i$ to a precision of 1.35\,km\,s$^{-1}$ for stars with $V \sin i$ $\geq$ 5\,km\,s$^{-1}$. However, measurements of $\log g$ from wavelet analysis are not reliable beyond confirming dwarf-like gravity ($\log g \approx 4.5$ dex). Subsequently, we fit the wings of the magnesium triplets with spectral synthesis by fixing $T_{\rm eff}$, [Fe/H] and $V\sin i$ and changing $\log g$ until an acceptable fit was found. 
 
\subsubsection{INT - synthesis}

INT observations of J1847$+$39 are unsuitable for wavelet analysis since there only a small wavelength region around the H$\alpha$ line was observed. The spectral synthesis technique was used to measure $T_{\rm eff}$ from the wings of the H$\alpha$ line and [Fe/H] from a limited number of Fe lines around the H$\alpha$ line. There are no gravity sensitive lines visible in the INT spectra and so we have not attempted to estimate $\log g$ from these spectra. 


\subsection{Ephemerides and first estimates for transit parameters}\label{ephem}

WASP photometry was used to obtain first estimates for the parameters of the transit and to determine a prior for the ephemerides; it was not used to determine the physical properties of the M-dwarf companions as transit depths can be unreliable. Using the framework of \cite{Beatty2007}, we obtain first order approximations to the ratio of semi-major axis, $a$, and the radius of the primary star, $R_{\star}$, using the width of the transit, $\Delta t_{\rm tr}$, 

\begin{equation}\label{approx_r_star}
\frac{R_{\star}}{a} \approx \pi \frac{\Delta t_{\rm tr}}{P},
\end{equation}
and the ratio of the radii, $k$, can be estimated from

\begin{equation}\label{approx_k}
k = \frac{R_{2}}{R_{\star}} \approx \sqrt{\Delta m},
\end{equation}
where $R_2$ is the radius of the M-dwarf companion and $\Delta m$ is the depth of transit caused by the M-dwarf. 
To measure $P$ and $T_0$, we use the method of \citet{1956BAN....12..327K} to compute accurately the epoch of minimum of each complete eclipse in the WASP photometry. Bayesian sampling was used to minimise the correlation between the uncertainties in $T_0$ and $P$. We generated 20,000 draws from 12 walkers and selected the reference period and time of minimum from the solution with the highest log-likliehood. The uncertainty for each parameter was estimated from the standard deviation of each parameters PPD and was used as a prior in the orbital fit (Sec. \ref{orbital_fit}). We inspected the difference between calculated models and observed epochs to concluded that there is no evidence of transit-timing variations for any of the five EBLMs.

\subsection{Rotational modulation}

Each system has thousands of observations from the WASP survey which have been taken over many years. Consequently, it is possible to measure variations in the light-curve caused by spot coverage or tidal interactions.  We used the method outlined in \citet{2011PASP..123..547M} to search the WASP photometry for frequencies attributed to rotational modulation. Each season of photometry is treated separately and in-transit data were excluded. We inspected the periodogram and false-alarm probabilities (FAP) for each system to assess the reliability of any detected periods. The false-alarm probabilities were calculated using the method of \citet{1989ApJ...338..277P}. We also phase-folded the light-curve at any detected period to check for ellipsoidal variation. An example periodogram for J0218$-$31 can be seen in Fig. \ref{WASP2308-46_period}. The Lomb-Scargle periodogram for each season for all targets can be found in Appendix \ref{appendix:lomb}.

\begin{figure*}[htb]
  \centering
  \includegraphics[width=\textwidth]{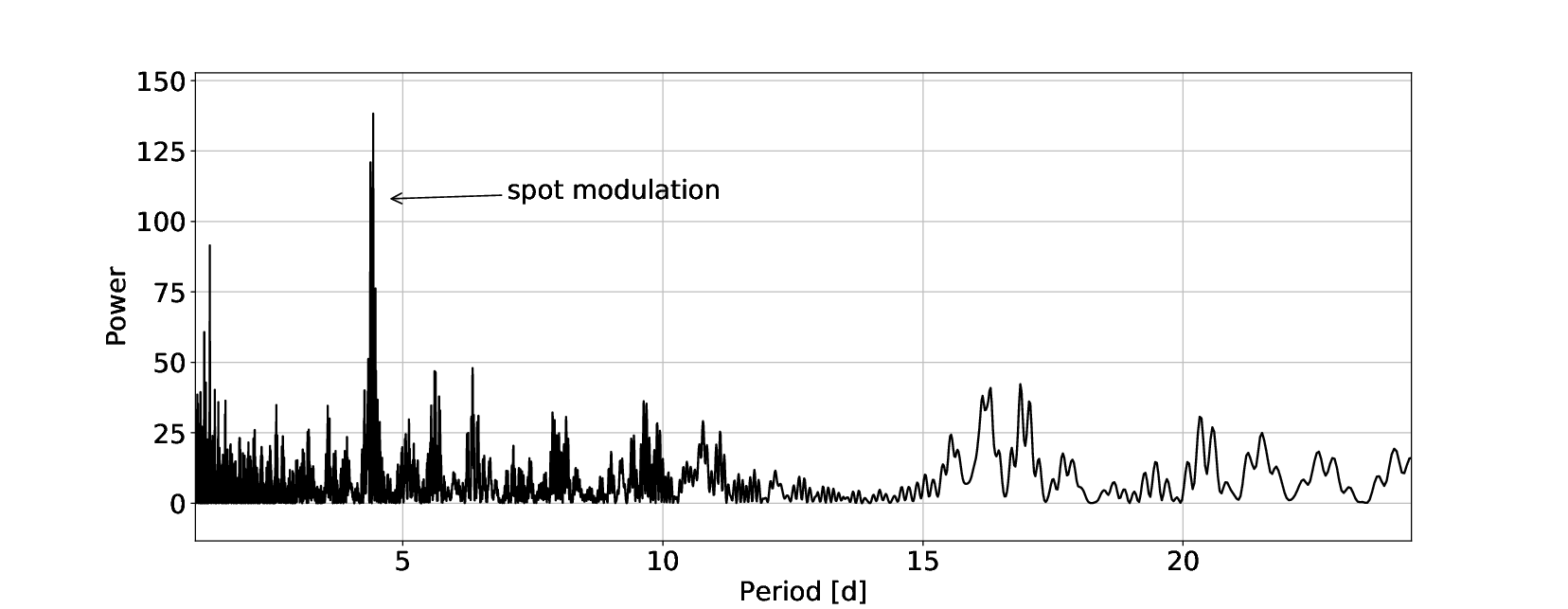}
  \caption{The power spectrum of J0218$-$31 using 3 seasons of WASP photometry. }
  \label{WASP2308-46_period}
\end{figure*}


\subsection{Combined radial velocity and light curve fit}\label{orbital_fit}


We fited all follow-up  photometry (from SAAO, CTIO and HAO) and radial velocity measurements simultaneously to obtain the final orbital solution for each system. We performed a $\chi^2$ fit in a Bayesian framework to estimate the PPD of each parameter in the vector model. The vector model of parameters includes photometric zero-points for each $i^{th}$ light-curve --$zp_i$, $R_{\star}/a$, $k$, the impact parameter --$b = a\cos(i)/R_{\star}$, $\rm T_0$, $P$, the limb-darkening temperature --$T_{\rm eff,ld}$, the semi-amplitude of radial velocity measurements --$K$, the systematic radial velocity -- $\gamma$ and the change in systematic radial velocity with time --$d(\gamma)/dt$. First estimates of $R_{\star}/a$, $k$, $\rm T_0$ and $P$ were obtained from photometry described in  Sec. \ref{ephem}. The first estimate of $T_{\rm eff,ld}$ was the spectroscopic value of $T_{\rm eff}$ from Sect. \ref{atmospheric_parameters}. Instead of fitting the argument of the periastron ($\omega$) and the eccentricity ($e$), we choose to use  $f_c = \sqrt{e} \cos \omega$ and  $f_s = \sqrt{e} \sin \omega$  since these have a uniform prior probability distribution and are not strongly correlated with each other. We also include a jitter term  ($\sigma_J$) to account for spot activity which can introduce noise in to the radial velocity measurements \citep{2006ApJ...642..505F}. We used $T_{\rm eff,ld}$ to interpolate coefficients for the Claret limb-darkening law (provided with the python package \textsc{ellc}, \citet{Maxted2016}) using fixed values of [Fe/H] and $\log g$ from Sect. \ref{atmospheric_parameters}. The stellar metalicity and surface gravity are fixed when interpolating limb-darkening coefficients as varying them has a second-order effect relative to the effective temperature. We used a Gaussian prior for $T_{\rm eff,ld}$ using the value of $T_{\rm eff}$ from Sect. \ref{atmospheric_parameters} with width equal to the uncertainty of $T_{\rm eff}$ measurements from the wavelet method, 85\,K. The follow-up photometry for each system is modelled with \textsc{ellc} assuming detached and spherical star-shapes. Gaussian priors for $ \rm T_{\rm 0}$ and $P$ from Sect. \ref{ephem} were used to constrain the ephemerides of each system. 

\begin{figure*}
    \centering
    \includegraphics[width=\textwidth]{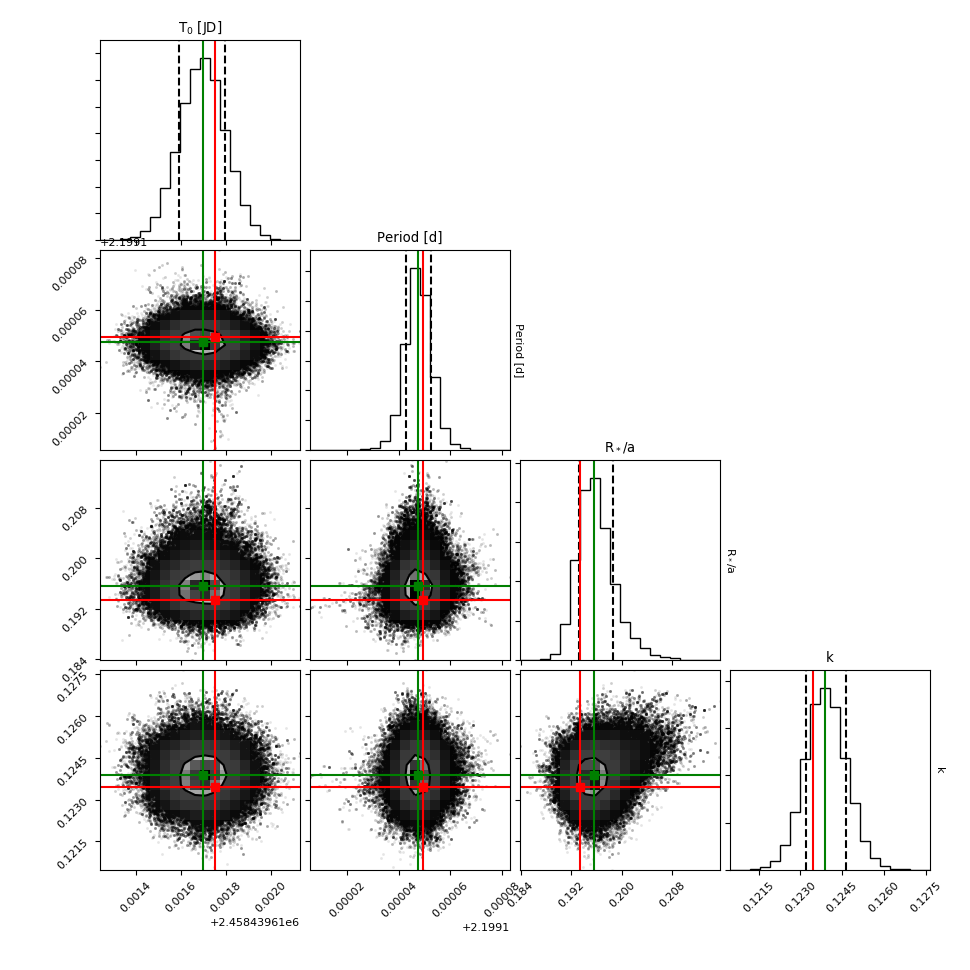}
    \caption{The orbital solution of J2349-32 with 1-$\sigma$ contours plotted. We show the median value of each parameter from the Markov chains (green) along with the the solution with the highest log-likliehood (red). }
    \label{fig:medianVSmax}
\end{figure*}

 We compare these models to data using a Bayesian framework with the likelihood function $\mathcal{L}(\textbf{d}|\textbf{m}) = \exp (-\chi^2/2)$, with

\begin{equation}\label{chi_squared}
\begin{split}
\chi^{2} = \sum_{i=1}^{N_{mag}} \frac{(m_{\rm i} - m_{\rm model})^2}{ \sigma_{m_{\rm i}}^2} &+ \sum_{i=1}^{N_{rv}} \frac{(rv_i - rv_{\rm model})^2}{\sigma_J^2 +  \sigma_{\rm rv_i}^2} \\ &+  \frac{(T_{\rm eff,ld} - T_{\rm eff})^2}{\sigma_{T_{\rm eff}}^2} .
\end{split}
\end{equation}
Here,  $m_{\rm i}$ and $rv_{\rm i}$ represent the $i^{th}$ measurement of magnitude and radial velocity with standard errors $\sigma_{m_{\rm i}}$ and $\sigma_{\rm rv_i}$,  respectively. We initiated 50 walkers and generated 50,000 draws, after an initial burn-in phase of 50,000 draws using \textsc{emcee}. We initially selected the model with the highest value of $\mathcal{L}(\textbf{d}|\textbf{m})$ from the PPD to extract the best-fitting model parameters. For J2308$-$46 and J1847$+$39 we found these values to be up to 1-$\sigma$ away from the median value of each parameters PPD (Fig. \ref{fig:medianVSmax}), and so we chose to use the median value from each parameters PPD instead. The uncertainties were calculated from the largest difference between the median and the $16^{th}$ and $84^{th}$ percentile of the cumulative PPD for each parameter from the second chain.

\subsubsection{Rossiter-McLaughlin}
We obtained radial velocity measurements of J0218$-$31 during the primary transit which display variations in radial velocity caused by the Rossiter-McLaughlin effect. Unlike \citet{Triaud2017} who excluded such measurements when determining the spectroscopic orbit, we fit these measurements simultaneously with out-of-transit radial velocities and follow-up transit photometry. The orbital fit for this system required two more de-correlated parameters,  $\sqrt{V \cos i} \sin \lambda$ and $\sqrt{V \sin i} \cos \lambda$, where $\lambda$ is sky-projected angle between the orbital and stellar rotation angular momentum vectors.

\subsubsection{Star shapes}
\begin{figure}[htb]
  \centering
  \includegraphics[width=0.5\textwidth]{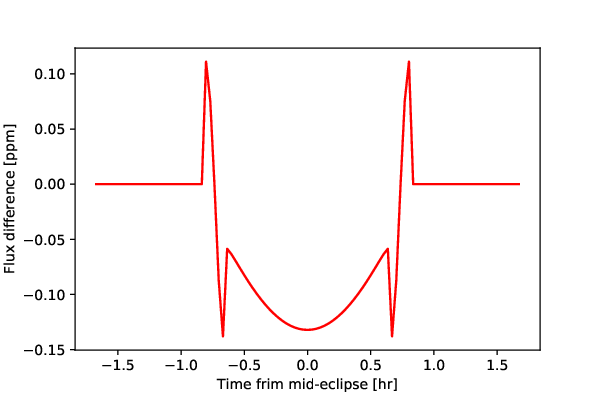}
  \caption {The difference between the spherical model and Roche model  of J2308$-$46 using \textsc{ellc}.}
  \label{model_diff}
\end{figure}

 We found ellipsoidal variations in the WASP photometry of J2308$-$46 which required the use of Roche geometry to estimate the initial  transit parameters from the WASP photometry. The follow-up photometry of J2308$-$46 was fitted using a spherical star shape  with the assumption that there is only a small amount of out-of-eclipse photometry, which was detrended. A caveat is that the spherical volume of the star will not be the same as the volume of the triaxial ellipsoid used to approximate its shape with \textsc{ellc}. We assessed the magnitude of this problem by comparing the models for J2308$-$46 where both stars are described by spheres to those where both stars are described using Roche models (Fig. \ref{model_diff}).  We find a maximum difference of $\approx 0.1\, \rm ppm$ which is far below the white-noise level (a few thousand ppm) and so we do not attempt to correct for this. The final orbital solution for all EBLMs assumes detached and spherical star-shapes and does not use Roche geometry.

\subsubsection{M-dwarf luminosity and transit depth}\label{M_dwarf_luminoisity}

\begin{table}
\caption{Modification of primary transit depths using \textsc{phoenix} model spectra. Uncertainties in $F_2 / F_*$ and  $\Delta$ Depth account for the uncertainties in stellar atmospheric parameters and $k$ (see Table \protect\ref{results_table}).} 
\label{table:transit_modification} 
\centering 
\begin{tabular}{c c c c} 
\hline\hline 
Target & Filter & $F_2 / F_*$ (\%) & $\Delta$ Depth [ppm] \\ 
\hline 
J2349$-$32 & I & $3.11 \pm 0.39$ & $55 \pm 17$  \\ 
J2308$-$46 & R & $2.86 \pm 0.34$ & $9 \pm 11$ \\ 
J0218$-$31 & g' & $6.33 \pm 0.22$ & $66 \pm 8$ \\ 
J0218$-$31 & r' & $6.35 \pm 0.22$ & $65 \pm 9$ \\ 
J0218$-$31 & i' & $6.35 \pm 0.22$ & $67 \pm 8$ \\ 
J0218$-$31 & z' & $6.33 \pm 0.22$ & $66 \pm 8$ \\ 

J1847$+$39 & CBB & $4.19 \pm 0.50$ & $348 \pm 45$  \\ 
J1847$+$39 & g' & $4.19 \pm 0.50$ & $348 \pm 46$ \\ 
J1847$+$39 & z' & $4.19 \pm 0.50$ &$348 \pm 46$ \\ 
J1436$-$13 & R & $7.47 \pm 0.45$ & $465 \pm 402$ \\ 
\hline 
\end{tabular}
\end{table}

The orbital solution assumes that the transit depth is not modified by the luminosity of the transiting M-dwarf. The justification of this assumption requires some foresight of the results (Table \ref{results_table}). The flux contribution depends on the transmission profile observations were made in and the stellar atmospheric parameters of each star. The modification of the primary transit depth further depends on the ratio of the radii, $k$. For the primary star, these quantities are estimated from spectral analysis. For the M-dwarf companion, we use $\log g_2$ from Table \ref{results_table} which was determined from the orbital solution and assume the metalicity is the identical to the host star under the premise they both formed in the same molecular cloud. The temperature of the M-dwarf companion was estimated from the MESA stellar evolution models \citep{2016ApJS..222....8D,2016ApJ...823..102C,2011ApJS..192....3P,2013ApJS..208....4P,2015ApJS..220...15P} using the nominal mass, age and composition. For each band-pass in each EBLM system, we used the following procedure:
\begin{enumerate}
    \item We interpolated high resolution \textsc{PHOENIX} model spectra \citep{2013A&A...553A...6H} for each star.
    
    \item The model spectrum for the host and the M-dwarf were convolved them with transmission profiles of the filters that their respective photometric transits were observed in. The ratio of the M-dwarf flux to the primary star flux ($F_2 / F_\star$) could then be calculated.
    
    \item The surface brightness ratio ($k^2 \times F_2/F_\star$) was calculated along with the change in primary transit depth accounting ($\Delta$ Depth).
\end{enumerate}
This process was repeated 1000 times for each transmission profile of each EBLM system by perturbing nominal measurements by their respective uncertainties in Table \ref{results_table}. The median value of $F_2/F_\star$ and $\Delta$ Depth were adopted as nominal measurements with uncertainties equal to the standard deviation of all draws (Table \ref{table:transit_modification}).  

EBLMs J2349$-$32, J2308$-$46 and J0218$-$31 have primary transit depths which are less than 100\,ppm shallower when accounting for the luminosity of M-dwarf companions. J1847$+$39  and J1436$-$13 have primary transits around 400\,ppm shallower due to the redder filters we obtained transit photometry with and higher values of $k$. The larger relative uncertainty for $\Delta$ Depth originates from a poorly constrained ratio of radii ($k$) due to a high impact parameter. We assume this effect is negligible and we do not correct for the light of the M-dwarf in this work.


\subsection{Mass and age estimates}

Breaking the degeneracy between the mass of the primary star -- $M_{\star}$, the M-dwarf companion -- $M_2$, and the age of the system -- $\tau$, is non-trivial. One approach by \cite{HebbCollier-CameronLoeilletEtAl2009} uses Kepler's equation to estimate the density of the primary,

\begin{equation}\label{density}
\rho_{\star} = \frac{3 \pi}{GP^2} \left( \frac{a}{R_{\star}} \right)^3 - \frac{3M_2}{4 \pi R_{\star}^3}
\end{equation}
and then combine it with measurements of $\rm T_{\rm eff}$ and [Fe/H] to interpolate between stellar models for $M_{\star}$ and $\tau$. Typically this is repeated with a better estimate of $M_{\star}$ until the solution converges iteratively. Another approach uses empirical mass and radius calibrations \citep{Southworth2011, Torres2013} to obtain $M_{\star}$ and $R_{\star}$.  These are combined with $k$ and eqn. \ref{density} to obtain $M_2$ and $R_2$. Yet another approach by \cite{TriaudHebbAndersonEtAl2013} is to mix the two methods while fitting alongside orbital parameters. The mass function \citep{Hilditch2001} can be expressed in terms of radial velocity parameters,

\begin{equation}\label{mass_function}
f(m)  =  \frac{(M_2 \sin i)^3}{(M_{\star} + M_2)^2} =  (1-e^2)^{\frac{3}{2}} \frac{P K^3}{2 \pi G},
\end{equation}
where $i$ is the inclination of the orbit and $G$ is the gravitational constant. The middle and right part of Eqn. \ref{mass_function} can be equated and solved numerically for $M_2$ assuming an initial guess of $M_{\star}$ from empirical calibrations. Stellar models are interpolated to give a new estimate of $M_{\star}$. The better value of $M_{\star}$ can be used to iteratively solve Eqn. \ref{mass_function} and generate better estimates of $M_{\star}$ and $M_2$ until a solution is converged upon \citep{TriaudHebbAndersonEtAl2013}. A final method relies on three assumptions: (1) the circularisation timescale ($\tau_{circ}$) is much shorter than $\tau$, (2) the rotation is synchronised ($\tau \gg \tau_{syn})$ and (3) that rotational and orbital inclination are the same. Under these assumptions it is possible to directly calculate the mass and radius of both components (see Eqns. 14-17 of \citealt{Beatty2007} or Eqns. 2-5 of \citealt{Zhou2014}). We used Eqns. 2-5 of \citet{Zhou2014} to estimate the masses and radii for the EBLMs presented in this work (Table \ref{results_table}). We found a significant discrepancy with our method likely arising from the lack of synchronicity in these systems. Furthermore, it is not enough to assume tidal circularisation and synchronisation from a short orbital period alone \citep{2019arXiv190305686F}.

To estimate the mass and age of the primary star we combined the atmospheric parameters (Sect. \ref{atmospheric_parameters}) and the best fitting orbital solution (Sect. \ref{orbital_fit}) and interpolate between evolutionary models computed with the \textsc{garstec} stellar evolution code \citep{2008Ap&SS.316...99W}. We make no assumptions regarding circularised or synchronised orbits. We used a modified version of the open-source code \textsc{bagemass} \citep{2015A&A...575A..36M} tailored exclusively for EBLM systems (\textsc{eblmmass}). \textsc{eblmmass} uses the jump parameters of age, primary mass ($M_{\star}$), the initial iron abundance in dex $\rm [Fe/H]_i$, $\rm M_2$ and the full-width half maximum of the transit $w$. The vector of  observed parameters is given by $\textbf{d} = (f(m),\,\rm T_{\rm eff}, \, \rm \log L_{\star}, \, \rm [Fe/H]_s, R_{\star}/a, w )$ where $\log L_{\star}$ is the luminosity of the primary star and $\rm [Fe/H]_s$ is the surface metal abundance in dex and $w$ is the transit width. The model parameters are $\textbf{m} = (M_{\star}, M_2, \tau, \rm [Fe/H]_i, w )$. $\rm [Fe/H]_s$ differs from the initial abundance ($\rm [Fe/H]_i$) due to diffusion and mixing processes throughout stellar evolution. The \textsc{garstec} evolutionary models used here are the same as the ones used in \textsc{bagemass}. \textsc{garstec} uses the FreeEOS\footnote{http://freeeos.sourceforge.net} equation of state \citep{2003ApJ...588..862C} and standard mixing length theory for convection \citep{1990sse..book.....K}. The mixing length parameter used to calculate the default model grid is $\alpha_{\rm MLT} = 1.78$. With this value of $\alpha_{\rm MLT}$ \textsc{garstec} reproduces the observed properties of the present day Sun assuming that the composition is that given by \citet{1998SSRv...85..161G}, the overall initial solar metallicity is $Z_{\sun} = 0.01826$, and the initial solar helium abundance is $Y_{\sun} = 0.26646$. These are slightly different to the value in \citet{2013PhRvD..87d3001S} because we have included additional mixing below the convective zone in order reduce the effect of gravitational settling and so to better match the properties of metal-poor stars. Due to the effects of microscopic diffusion, the initial solar composition corresponds to an initial iron abundance [Fe/H]$_{\rm i} = +0.06$ dex. The stellar model grid covers the mass range $0.6 M_{\sun}$ to $2.0 M_{\sun}$ in steps of $0.02 M_{\sun}$. The grid of initial metallicity values covers the range [Fe/H]$_i$ = $-0.75$ dex to $-0.05$ dex in steps of $0.1$ dex and the range [Fe/H]$_i$ = $-0.05$ to +$0.55$ in $0.05$ dex steps. 

To obtain $M_2$ from $f(m)$, $M_{\star}$ and $P$, we need to know inclination from the transit light-curve. Degeneracies between $i$, $R_{\star}/a$ and $k$ are such that we choose to fit the full-width half maximum of the transit,
\begin{equation}\label{fwhm_transit}
w = \frac{R_{\star}}{a} \frac{\sqrt{1-b^2}}{\pi}. 
\end{equation}
instead of the inclination. We implement a Gaussian prior on $\rm [Fe/H]_s$ from spectroscopy and use a uniform priors for age, $M_{\star}$ and $M_2$. We ran a burn-in chain of 100,000 draws before drawing an additional 50,000 draws to sample the PPD for $M_{\star}$, $M_2$ and $\tau$. The number of post-burn-in draws matches that of the orbital fit.


We use an up-to-date constant from IAU resolution B3 \citep{Mamajek2015} to calculate $a$ from $P$, $M_{\star}$ and $M_2$,
\begin{equation}
a = 4.208278 \times P^{\frac{2}{3}} (M_{\star} + M_2)^{\frac{1}{3}}.
\end{equation}
This can then be combined with $R_{\star}/a$ and $k$ to calculate the PPD for $R_{\star}$ and $R_2$. We selected the the median value of each parameters PPD as our measurements, with uncertainty equal to the largest difference between the median and the $16^{th}$ and $84^{th}$ percentile of the cumulative PPD for each parameter from the second chain.

\begin{figure*}[htb]
  \centering
  \includegraphics[width=\textwidth]{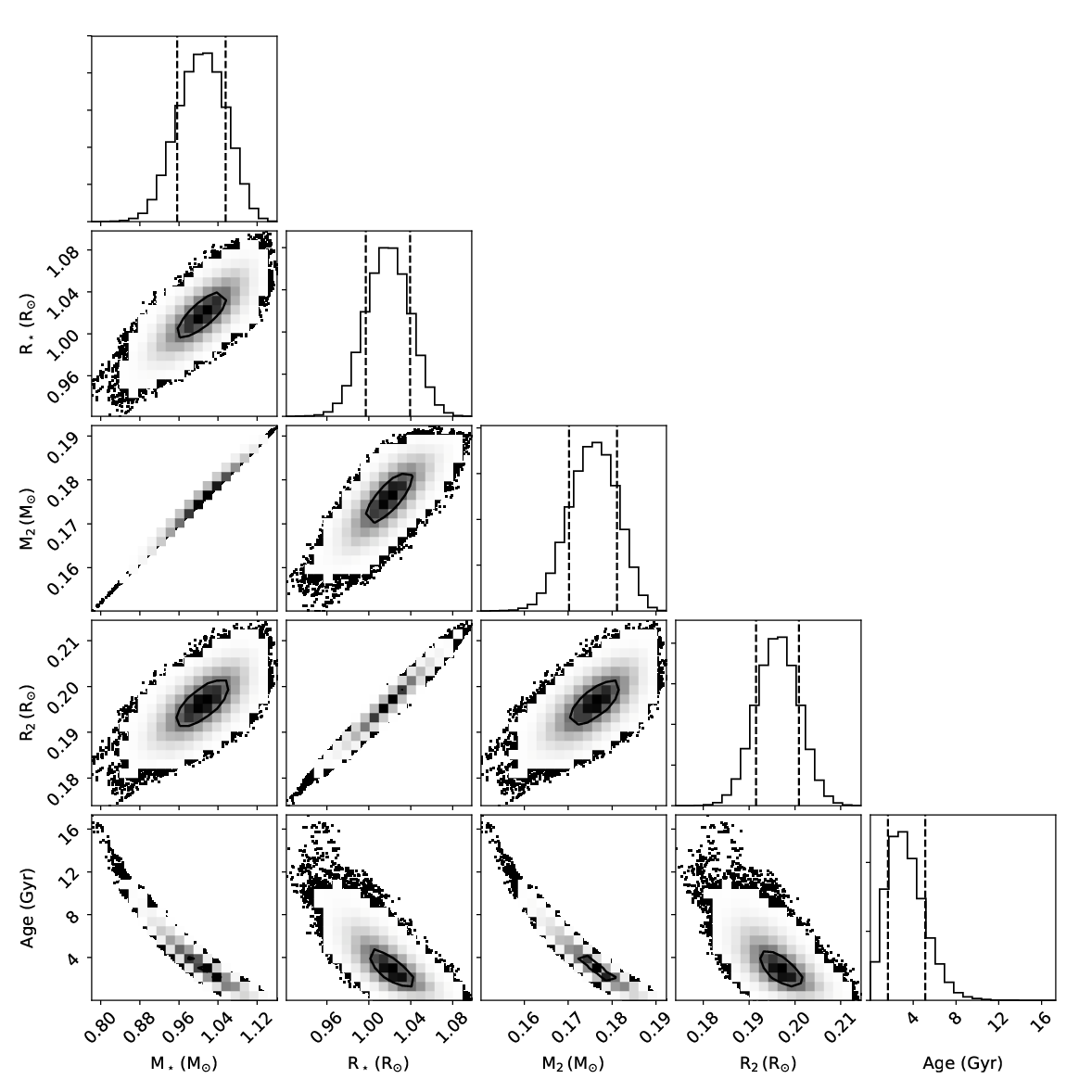}
  \caption{The posterior probability distributions from  \textsc{eblmmass} for J2349$-$32. Over plotted are the 1-$\sigma$ contours.}
  \label{WASP2349-32_EBLMMASS}
\end{figure*}


\section{Results}\label{results}

\begin{table*}
\caption{A description of 5 EBLM systems measured in this work. This table gives an overview of key results from SED fitting, spectroscopy, orbital fitting and \textsc{eblmmass}. For J2308-46 and J0218-31 we report both solutions for mass and age. }              
\label{results_table}      
\centering   
\resizebox{0.9\linewidth}{!}{
\begin{tabular}{l c c c c c}          
\hline\hline                        
 & J2349$-$32 
 & J2308$-$46 
 & J0218$-$31 
 & J1847$+$39 
 & J1436$-$13 \\
\hline 

\multicolumn{3}{l}{From SED fitting} \\
$\rm T_{\rm eff, phot}$ (K) 
& $6090 \pm 90$ 
& $6270 \pm 140$ 
& $6020 \pm 100$ 
& $6210 \pm 220$ 
& $6080 \pm 360$ \\

$\rm E(B-V)$ 
& $0.017 \pm 0.017$ 
& $0.032 \pm  0.022$ 
& $0.030 \pm 0.020$ 
& $0.073 \pm 0.042$ 
& $0.031 \pm  0.024 $ \\

$g'_0$   
& $11.708 \pm 0.067$
& $11.565 \pm 0.092$
& $10.045 \pm 0.082$ 
& $11.753 \pm 0.167$ 
& $12.502 \pm 0.121$ \\ \\

\multicolumn{3}{l}{From spectroscopy} \\
$\rm T_{\rm eff}$ $\rm (K)$    
& $6130 \pm 85$     
& $6185 \pm 85$     
& $6100 \pm 85$    
& $6200 \pm 85$   
& $6310 \pm 85$\\

$\log g$ (dex)           
& $4.42 \pm 0.13$    
& $4.21 \pm 0.13$  
& $4.05 \pm 0.13$   
& $4.44 \pm 0.13$  
& $4.25 \pm 0.13$\\

$\xi_t \, (\rm km\,s^{-1})$ 
& $1.05 \pm 1.50$ 
& $1.07 \pm 1.50$ 
& $1.03 \pm 1.50$ 
& $1.08 \pm 1.50$ 
& $1.14 \pm 1.50$ \\

$v_{mac}\, (\rm km \, s^{-1})$  
& $4.23 \pm 1.50$ 
& $4.95 \pm 1.50$ 
& $4.94 \pm 1.50$ 
& $4.55 \pm 1.50$ 
& $5.41 \pm 1.50$ \\

Vsin$i$ (km\,s$^{-1}$) 
& $11.50 \pm 1.35$ 
& $39.83 \pm 1.35$ 
& $9.00 \pm 1.35$ 
& $10.00 \pm 1.35$  
& $18.80 \pm 1.35$ \\

$\rm [Fe/H]$ (dex)  
&  $-0.28\,\pm\,0.06$ 
&  $-0.15\,\pm\,0.06$ 
&  $0.15 \pm 0.06$ 
& $-0.25 \pm 0.06$ 
&  $-0.10 \pm 0.06$ \\

$\log \rm A(Li)$ + 12 
& $2.4 \pm 0.1$ 
& - 
& $3.1 \pm 0.1$ 
& - 
& - \\

\\
From orbital fit \\
$R_{\star}/a$  &
$0.0980 \pm 0.0003$ & 
$0.1934 \pm 0.0030$ &
$0.0988 \pm 0.0029$ &
$0.0570 \pm 0.0005$ &
$0.1084 \pm 0.0005$ \\

$R_{2}/a$  & 
$0.0188 \pm 0.0003$ &
$0.0239 \pm 0.0001$ &
$0.0165 \pm 0.0006$ &
$0.0162 \pm 0.0002$ &
$0.0290 \pm 0.0040$ \\

$k$  &  
$0.1923 \pm 0.0002$ &
$0.1234 \pm 0.0007$ &
$0.1685 \pm 0.0033$ &
$0.2842 \pm 0.0010$ &
$0.2841 \pm 0.0403$ \\

$b$  &  
$0.33\pm 0.01$ &
$0.05\pm 0.08$ &
$0.72\pm 0.02 $ &
$0.41\pm 0.02 $ &
$0.87\pm 0.07 $ \\

$T_{\rm eff, \, ld}\, \rm (K)$ &
$6105 \pm 260$ & 
$6530 \pm 320$ &
$6109 \pm 400$ &
$6860 \pm 260$ &
$6072 \pm 360$ \\

\\
$K\, \rm (km\,s^{-1})$ &
$21.92 \pm 0.02$ &
$23.70 \pm 0.17$ &
$27.80 \pm 0.01$ &
$27.69 \pm 0.83$ &
$46.50 \pm 0.07$ \\

$f_s$ &
$0.003 \pm 0.023 $ & 
$-0.003 \pm 0.050 $ &
$-0.008 \pm 0.051$ &
$0.070 \pm 0.052 $ &
$0.022 \pm 0.052 $ \\

$f_c$ & 
$0.037 \pm 0.027$ &
$0.104 \pm 0.061$ &
$-0.001 \pm 0.050$ &
$-0.451 \pm 0.013 $ &
$0.032 \pm 0.027 $ \\

$e$ & 
$0.001 \pm 0.002$ & 
$0.009 \pm 0.011$  &
$\leq 0.001$ &
$0.209 \pm 0.014$ &
$0.002 \pm 0.002$ \\

$\omega \, (^{\circ})$ & 
$90 \pm 40$ &
$269 \pm 33$  &
- &
$351 \pm 18$  &
$34 \pm 24$  \\

$\gamma \,\rm (km\,s^{-1})$ & 
$1.660 \pm 0.053$ &
$6.073 \pm 0.831$ &
$48.640 \pm 0.010$ &
$-67.431 \pm 0.527$ &
$6.718 \pm 0.257$ \\

$d(\gamma)/dt \,  \rm (ms^{-1} \,yr^{-1})$ & 
$4.2 \pm 3.59$ & 
$0.8 \pm 0.3$ &
$-69.9 \pm 4.1$ &
$-71.9 \pm 21.7$ &
$-23.5 \pm 86.1$ \\

$\sqrt{V\sin i}\sin \lambda$ &
- &
- &
$0.131 \pm 0.385$ &
- &
- \\

$\sqrt{V\sin i}\cos \lambda$ &
- &
- &
$3.204 \pm 0.331$ &
- &
- \\

\\
$T_0\, \rm(HJD_{TDB})$ &
\begin{tabular}{@{}c@{}}$2454215.89924$ \\ $\pm 0.00007$\end{tabular} &
\begin{tabular}{@{}c@{}}$2458439.61178$ \\ $\pm 0.00010$\end{tabular} &
\begin{tabular}{@{}c@{}}$2455613.39961$ \\ $\pm 0.00005$\end{tabular} &
\begin{tabular}{@{}c@{}}$2454234.68992$ \\ $\pm 0.00010$\end{tabular} &
\begin{tabular}{@{}c@{}}$2454625.48423$ \\ $\pm 0.00008$\end{tabular} \\

$P \, \rm (d)$ &
\begin{tabular}{@{}c@{}}$3.5496972$ \\ $\pm 0.0000027$\end{tabular} &
\begin{tabular}{@{}c@{}}$2.199187$ \\ $\pm 0.0000022$\end{tabular} &
\begin{tabular}{@{}c@{}}$8.884102$ \\ $\pm 0.0000111$\end{tabular} &
\begin{tabular}{@{}c@{}}$7.325177$ \\ $\pm 0.000003$\end{tabular} &
\begin{tabular}{@{}c@{}}$3.9975234$ \\ $\pm 0.000004$\end{tabular} \\ \\

\multicolumn{3}{l}{Assuming circularization and synchronization \protect\citep{Zhou2014}}\\
$\rm M_{\star}\,(\rm M_{\sun})$  &
0.48 &
1.76 &
0.49 &
3.468 &
1.60 \\

$\rm R_{\star}\,(\rm R_{\sun})$  &
0.80 &
1.73 &
1.58 &
1.44 &
1.49 \\

$\rm M_{2}\,(\rm M_{\sun})$ &
0.11 & 
0.22 &
0.21 &
0.63 &
0.58 \\

$\rm R_{2}\,(\rm R_{\sun})$ &
0.16 & 
0.21 &
0.27 &
0.41 &
0.42 \\

\\
\multicolumn{3}{l}{from \textsc{eblmmass}} \\
$\rm M_{\star}\,(\rm M_{\sun})$ 
& $0.991 \pm 0.049$ 
& \begin{tabular}{@{}c@{}}$1.223 \pm 0.049$ \\ $1.089  \pm 0.049$\end{tabular} 
& \begin{tabular}{@{}c@{}}$1.550 \pm 0.050$ \\ $1.340  \pm 0.050$\end{tabular} 
& $1.054 \pm 0.058$ & $1.185 \pm 0.073$  \\


$\rm R_{\star}\,(\rm R_{\sun})$ 
& $0.965 \pm 0.022$ 
& $1.534 \pm 0.041$ 
& $2.131 \pm 0.088$
& $1.003 \pm 0.0194$ 
& $1.360 \pm 0.063$ \\

$\rm M_{2}\,(\rm M_{\sun})$     
& $0.174 \pm 0.006$ 
& \begin{tabular}{@{}c@{}}$0.172 \pm 0.004$ \\ $0.182 \pm 0.005$\end{tabular}
& \begin{tabular}{@{}c@{}}$0.390 \pm 0.009$ \\ $0.427 \pm 0.009$\end{tabular}
& $0.303 \pm 0.014$ 
& $0.490 \pm 0.018$ \\


$\rm R_{2}\,(\rm R_{\sun})$     &
$0.202 \pm 0.005$ &
$0.189 \pm 0.005$ &
$0.361 \pm 0.020$ &
$0.287 \pm 0.006$ &
$0.408 \pm 0.061$ \\


$\rm Age\, \rm (Gyr)$           
& $2.3 \pm 2.0$ 
& \begin{tabular}{@{}c@{}}$3.8 \pm 0.6$ \\ $5.9\pm 1.1$\end{tabular}
& \begin{tabular}{@{}c@{}}$2.4 \pm 0.3$ \\ $3.8 \pm 0.4$\end{tabular}
& $1.1 \pm 1.8$
& $2.3 \pm 0.1$ \\

\hline                                             
\end{tabular}}
\end{table*}



\subsection{EBLM J2349$-$32}

\begin{figure}[htb]

  \centering
  \includegraphics[width=\textwidth/2]{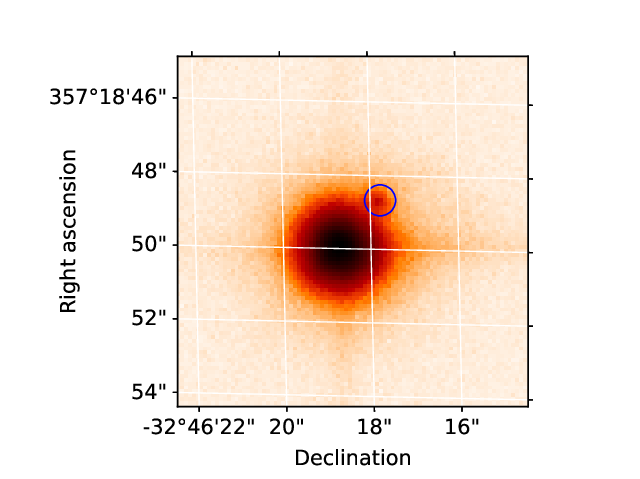}
  \caption{Lucky imaging of J2349$-$32 (red arm) revealing a close companion 1.3" (blue circle) away at a position angle of $308.6 \pm 0.6^\circ$ (blue circle).}
  \label{WASP2349-32_lucky}
\end{figure}

\begin{figure}[htb]
  \centering
  \includegraphics[width=\textwidth/2]{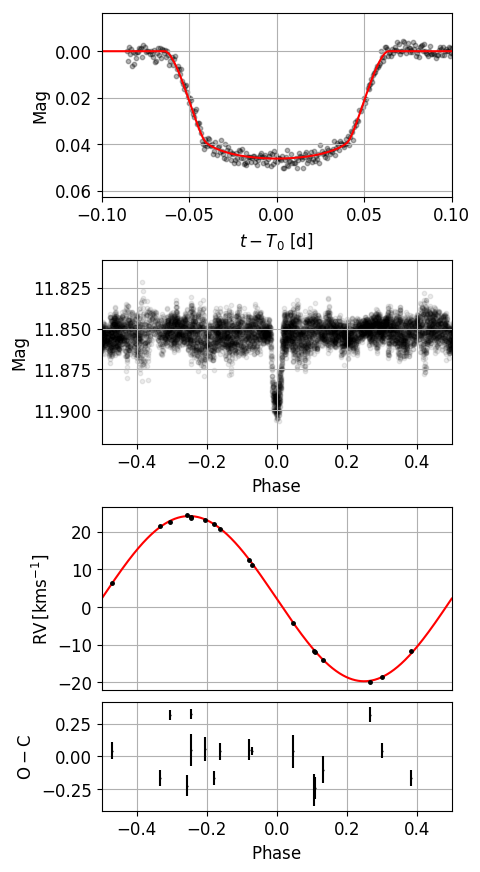}
  \caption{Orbital fit of J2349$-$32. (top panel) The detrended I-band light-curve from the SAAO 1-m telescope (black) with the best fitting transit model (red). (upper-middle panel) The phase-folded WASP lightcurve (black).  (lower-middle panel) Drift-corrected radial velocity measurements (black) with the best model (red). (bottom panel) The residuals from radial velocity model measurements.}
  \label{WASP2349-32-orbit}
\end{figure}

J2349$-$32 was observed over three consecutive years by the WASP project. In each season we found quasi-periodic signals at periods of $4.42\, \rm d$, $4.35\, \rm d$ and $4.42\, \rm d$ with amplitudes between 3--4\,mmag. Each of these signals has a false-alarm probability $<10^{-5}$ and so we assumed this is detection of the rotational period of the primary star ($P_{\rm rot} = 4.40 \pm 0.03$\,d). From the WASP photometry, we measured $\Delta t_{\rm tr} = 0.126$\,d and $\Delta m = 0.043$ mag. corresponding to $R_\star / a \approx 0.11 $ and $k \approx 0.21$.

The best SED fit ($\chi_{\rm red}^2 = 1.24$) corresponds to a star with spectral type F9 with a low reddening ($\rm E(B \rm- \rm V) \leq 0.034$ to 1-$\sigma$). This system was included in Gaia DR2 (source ID 2314099177602409856). The $G$  magnitude was measured to be $11.413$ and the parallax is $3.769\pm 0.092\, \rm mas$ ($265.32 \pm 6.47$\,pc).  Gaia DR2 shows a single star ($G = 15.219$) 48" away at a position angle of $111^\circ$ (source ID 2314099173307737088). This source is not included in the sky annulus of the SAAO 1-m photometry, but falls within the WASP aperture where it will contribute around 3\% of the total flux. The proper motions of this star and J2349$-$32 are significantly different in right ascension and declination so we concluded that they are not associated.

J2349$-$32 was observed with lucky imaging on 2017-07-08 where two companion stars were detected. A close companion was found at a separation of $1.402 \pm 0.013$" and position angle of $308.6^{\circ} \pm 0.6^{\circ}$ (Fig. \ref{WASP2349-32_lucky}). We measured the companion to be $5.55 \pm 0.08$\,magnitudes fainter in the TCI red-arm images; the companion was not sufficiently resolved in the TCI visible-arm images to obtain any reliable measurements. A second, distant companion was detected at a separation of $25.70 \pm 0.07$", position angle of $218.6 \pm 0.3^\circ$. We find that it is $9.0\pm0.3$\,magnitudes fainter with the TCI red-arm images and $8.5 \pm 0.3$\,magnitudes fainter in the visible-arm images. This is the same source identified by Gaia DR2 (source ID 2314099173307737088). If the closest companion is blended in the CORALIE and SAAO 1-m apertures, we estimate that it only contributes 0.6\,\% of the light and is too faint to significantly modify the transit light-curve. 

The 18 CORALIE spectra were combined to produce a spectrum with S/N$=40$. The analysis of this spectrum shows that the primary is a slightly metal-deficient star with a temperature consistent with the SED fit. There is a weak Li\,I line at 670.7\,nm from which we measured a lithium abundance $\log \rm A_{\rm Li}+12  = 2.4\,\pm \, 0.08$. This value was estimated by synthesising a small region around this line in \textsc{ispec} using fixed atmospheric parameters from wavelet analysis (Table \ref{results_table}) and manually adjusting the lithium abundance to obtain the best fit by-eye. 

The RVs were fitted simultaneously with a single transit in I-band from the SAAO 1-m telescope to obtain the best fitting orbital solution ($\chi_{\rm red}^2 = 0.93$; Fig. \ref{WASP2349-32-orbit}). The PPD for eccentricity is consistent with a circular orbit ($e \leq 0.05$ to $5$-$\sigma$). We find a negligible drift in systematic velocity ($\leq$ 15\,m\,s$^{-1}$\,yr$^{-1}$ to $1$-$\sigma$). The best-fitting limb-darkening temperature agrees with effective temperatures measured with SED fitting and wavelet analysis to better than 1-$\sigma$.


\textsc{eblmmass} predicts a primary star which has a mass and radius similar to the Sun, but is approximately 350\,K hotter. This is partly due to this being a metal-poor star, but also because it is approximately half the age of the Sun. The youthfulness of this star in conjunction with a convection zone which is unable to transport lithium to the core where it would be burnt may explain why lithium is detected with spectroscopy. The secondary component's mass is consistent with that of an M-dwarf below the fully convective limit.


\subsection{EBLM J2308$-$46}\label{results:J2308}


\begin{figure}[htb]

  \centering
  \includegraphics[width=\textwidth/2]{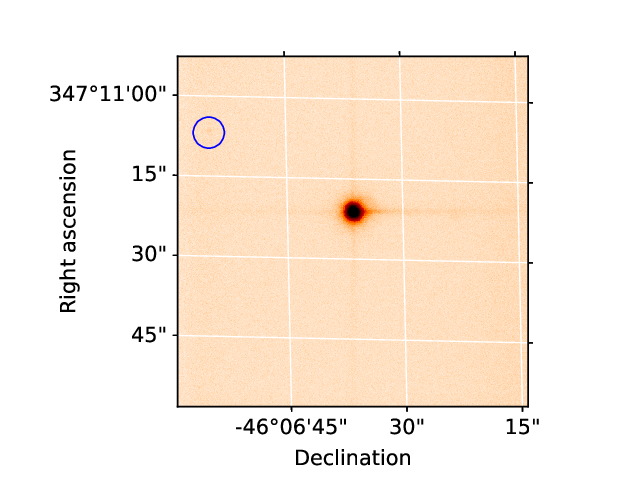}
  \caption{Lucky imaging of J2308$-$46 (red arm) revealing a companion 20" away with a position angle of 208$^\circ$ (blue circle).}
  \label{WASP2308-46_lucky}
\end{figure}

\begin{figure*}[htb]
  \centering
  \includegraphics[width=\textwidth]{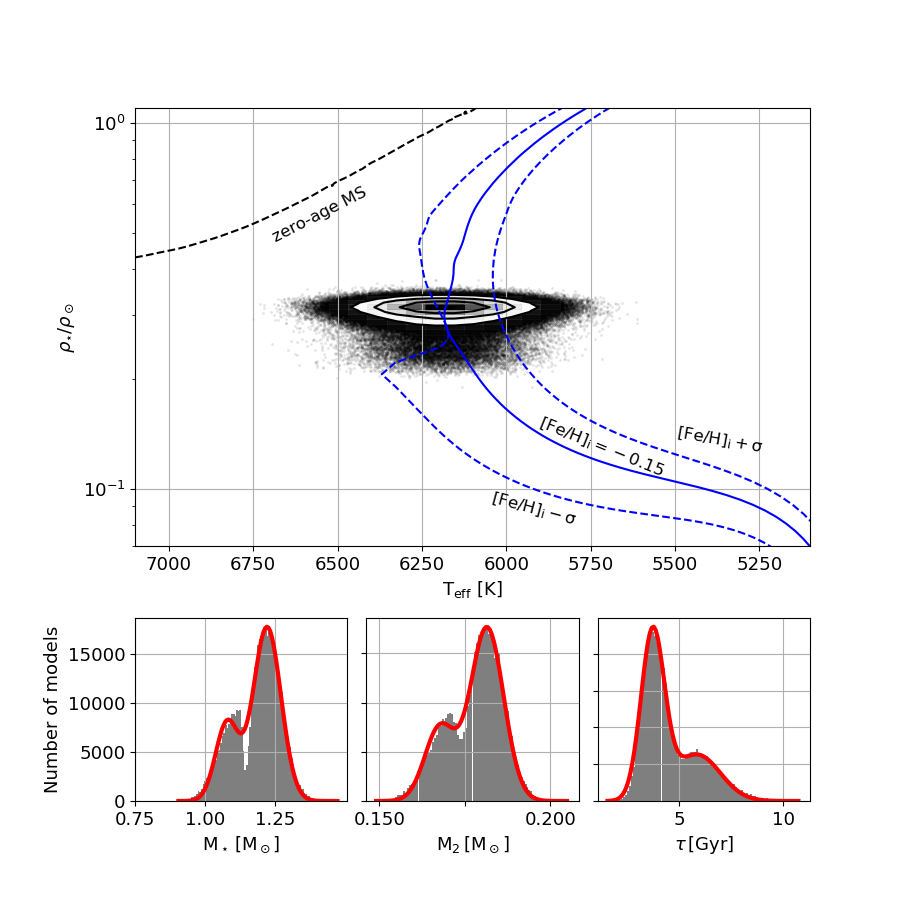}
  \caption{The PPD for the density and temperature of the primary star in J2308$-$46 is shown in the top panel. The zero-age main sequence is show(black-dashed) along with the best fitting isochrone (blue-solid) and the respective isochrones for $\pm 1$-$\sigma$ in [Fe/H]. The lower panels show the PPD distributions for $M_1$, $M_2$ and $\tau$ with best-fitting double-Gaussian models in red. }
  \label{WASP2308-46_HR}
\end{figure*}

\begin{figure}[htb]

  \centering
  \includegraphics[width=\textwidth/2]{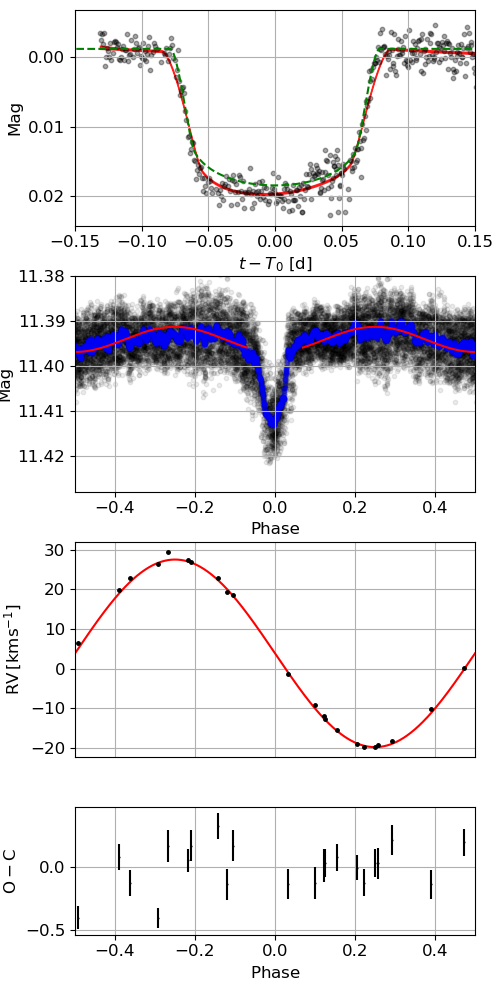}
  \caption{Orbital fit of J2308$-$46. (upper panel) R-band transit obtained from the SAAO 1-m telescope (black) with the best fitting transit model (green dashed). We plot the best fitting transit model generated using Gaussian processes (red). (middle panel) Phase-folded WASP observations (black) and observations binned into groups of 50 (blue). We also plot the Roche model used to approximate the out-of-transit photometry used to measure transit parameters from WASP photometry (red).  (lower panel) Drift-corrected radial velocity measurements (black) with the best fitting model (red) and residuals from the best fitting orbital model.}
  \label{WASP2308-orbit}
\end{figure}

J2308$-$46 has WASP photometry spanning 5 years. The last season of data had less than 400 data points so was excluded. We measured a strong $P/2$ signal in two seasons of data. Phase folding WASP photometry at this period reveals a moderate ellipsoidal variation with an amplitude of 5\,mmag (Fig. \ref{WASP2308-orbit}). We fixed parameters associated with ellipsoidal variation to produce a good out-of-transit fit to the WASP photometry ($q = M_{2}/M_{\star}=0.05$, gravity darkening coefficient = 0.1) to measure $\Delta t_{\rm tr} = 0.109$\,d and $\Delta m = 0.018$\,magnitudes and estimate the transit parameters $R_\star / a \approx 0.20$ and $k\approx0.13$.

SED fitting measured the effective temperature of the primary star to be consistent with a spectral type F7 ($T_{\rm eff} = 6270 \pm 140$\,K; $\chi^2_{\rm red} = 0.77$) with a low reddening ($E(B-V) \leq 0.054$ to 1-$\sigma$). This system is included in the Gaia DR2 catalogue (Source ID 6539811294185397120; $G = 11.361$) with a parallax of $2.187 \pm 0.113$\,mas ($457.24 \pm  23.63$\,pc). There is a close companion  22.5" from J2308$-$46 at a position angle of $282^\circ$ ($G=15.388$; source ID 6539811500344886016). This is clearly resolved in the follow-up 1-m R-band photometry from SAAO and does not contaminate the sky annulus. It does fall within the WASP annulus, contributing approximately 3\% of the total flux. There is another source 48" away from J2308$-$46 at a position angle of $23^\circ$. This is also included in Gaia DR2 ($G = 14.919$; source ID 6539817204061452544) which is on the limits of the WASP aperture and would contribute less flux than the source at position angle $282^\circ$. J2308$-$46 was observed with lucky imaging on 7 July 2017 with only a single, faint companion being found, located $21.38 \pm 0.04$" away at a position angle of $208.2\pm0.4^\circ$ (Fig. \ref{WASP2308-46_lucky}). We measured magnitude differences of $8.2 \pm 0.3$\,magnitudes in the TCI red-arm images and $8.0\pm0.2$\,magnitudes in the TCI visible-arm images. This object is included in the Gaia DR2 catalogue with Source ID 6539811289890737408 with $G=19.538$. We compared the proper motion of J2308$-$46 
with this object
 and conclude they are not physically associated.

A total of 22 CORALIE spectra were co-added to produce a spectrum with S/N=$20$. Wavelet decomposition shows that the primary star is moderately-rotating (V$\sin i \, \approx 39 \, \rm km\,s^{-1}$) and metal poor ([Fe/H] = $-0.15$ dex). The primary star's effective temperature appears to be close to the temperature at which the Kraft break \citep{Kraft1967} becomes apparent. The Kraft break is an abrupt reduction in surface rotation for stars with effective temperatures below $\sim$6200\,K. This is due to the presence of a efficient magnetic dynamo which transfers angular momentum from the star through stellar winds resulting in magnetic breaking.  
 
 Fitting the follow-up photometry jointly with radial velocity measurements was non-trivial as clear systematic errors remained in the SAAO 1-m light-curve after initial detrending. We obtained an orbital solution in the same framework as EBLM J2349$-$32 but found an unacceptable fit around contact point 2 and the continuum prior to contact point 1 in the SAAO 1-m light-curve (see top panel of  Fig. \ref{WASP2308-orbit}). We attempted further detrending of the follow-up light-curve with airmass, CCD position and time but this did not successfully remove the problem. Instead, we decided to generate a red-noise model using Gaussian processes. We used the \textsc{celerite} package \citep{2017AJ....154..220F} and the following kernel with the default value of $\epsilon=0.01$ to approximate the Mat\'{e}rn-3/2 covariance function: 
\[ k(\tau_k) = \sigma^2\,\left[
  \left(1+1/\epsilon\right)\,e^{-(1-\epsilon)\sqrt{3}\,\tau_k/\rho}
  \left(1-1/\epsilon\right)\,e^{-(1+\epsilon)\sqrt{3}\,\tau_k/\rho} \right]. \]
 Here, $\tau_k$ is the time difference between two observations, $\rho$ is a parameter that controls the time scale over which observational errors are correlated and $\sigma$ controls the amplitude of such variations. The free parameters, $\log \rho$ and $\log \sigma$, tended to a value that over-fitted the noise in the light-curve if it remained as a free parameter in the joint fit. Instead, we adjusted these values ``by-eye'' until we found an acceptable red-noise model that accounted for the data around the second contact point ($\log \rho=2$ and  $\log \sigma=2$). The parameters were then fixed at these values to find an acceptable orbital solution ($\chi_{\rm red}^2 = 1.32$) in the same way as J2349$-$32.
 
The primary star is close the the ``blue-hook'' phase of its post main-sequence evolution (Fig. \ref{WASP2308-46_HR}). This results in two peaks in the PPDs for $M_\star$ , $M_2$ and $\tau$ which are consistent to within 2-$\sigma$. Both solutions could be valid and so it is a requirement to fit these systems to assess the likelihood and validity of each. We fitted double-Gaussian models to the PPDs of $M_\star$, $M_2$, $\tau$ and $a$ which have been sorted into 100 equal bins. We used the Levenberg-Marquardt algorithm to find the optimal model vector $\textbf{m} = (A_1, \mu_1, \sigma_1, A_2, \mu_2, \sigma_2)$ for the double Gaussian model:
\begin{equation}
    y = A_1e^{-\frac{(x-\mu_1)^2}{2\sigma_1^2}} + A_2e^{-\frac{(x-\mu_2)^2}{2\sigma_2^2}},
\end{equation}
where $x$ is the position of the bin, $y$ is the number of models in the respective bin, $\mu$ is the measurement of the model, $\sigma$ is the uncertainty associated with the model, and $A$ represents the number of models at the peak of the of the distribution. 
The resulting fit for J2308$-$46 isn't entirely satisfactory; the fitted values of $\mu$ do not entirely match up with the peaks of the PPD for $M_\star$ , $M_2$ and $\tau$. This is partly due to the PPDs being poorly described by a Gaussian. Other EBLMs (e.g. J0218$-$31) have double-peaked PPDs which are well described by Gaussian, so we decided to add additional uncertainty rather than seeking a more complex model. To account for this, we added an additional uncertainty of 2\% for $M_\star$ , $M_2$ and $\tau$ which was estimated by measuring the offset between the fitted values of $\mu$ and the peaks of the respective PPDs. Moreover, the width of each PPD ($\sigma$) is underestimated upon visual inspection leading to an additional 1\% uncertainty which was determined ``by-eye''. The total additional uncertainty for each $\sigma$ is 3\%. We assessed each solution using the ratio of likelihoods. We found that the younger solution ($\tau = 3.98 \pm 0.86$ Gyr) is preferred over the older solution ($\tau = 5.81 \pm 1.0$ Gyr) with a factor $\mathcal{L}(3.98\, \rm Gyr) / \mathcal{L}(5.81\, \rm Gyr)$ $\approx 3.07$. This is moderate evidence to favour the younger solution but far from conclusive so we report both solutions in Table \ref{results_table}.  


\subsection{EBLM J0218$-$31}


\begin{figure}[htb]
  \centering
  \includegraphics[width=\textwidth/2]{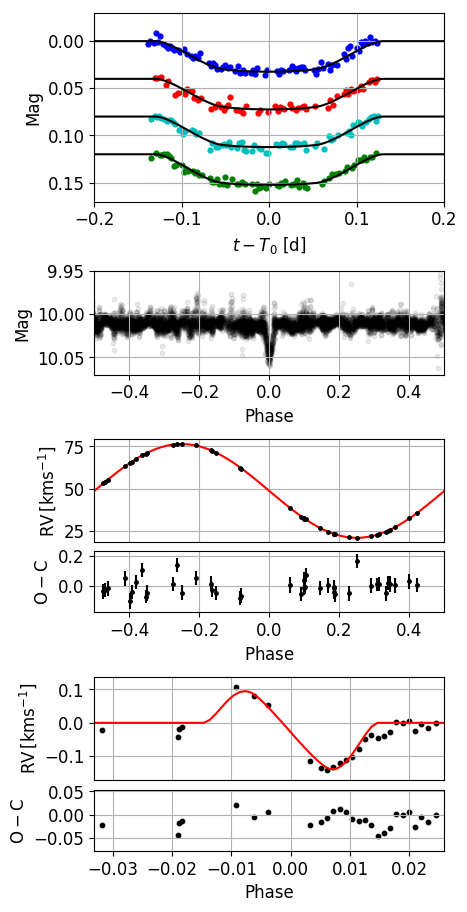}
  \caption{Orbital fit for J0218$-$31. (top panel) Transit photometry from CTIO in $g'$ (blue), $r'$ (red), $i'$ (cyan) and $z'$ (green) with best fitting models shown in black. (upper-middle panel) The phase-folded WASP lightcurve. (lower-middle panel) Drift-corrected radial velocity measurements from CORALIE with best fitting model plotted in red, along with residuals. (bottom panel) Drift-corrected radial velocity measurements during transit (the Rossiter–McLaughlin effect; black) with the best fitting model (red). Error bars have been omitted for clarity.}
  \label{WASP0218-31-orbit}
\end{figure}

\begin{figure*}[htb]
  \centering
  \includegraphics[width=\textwidth]{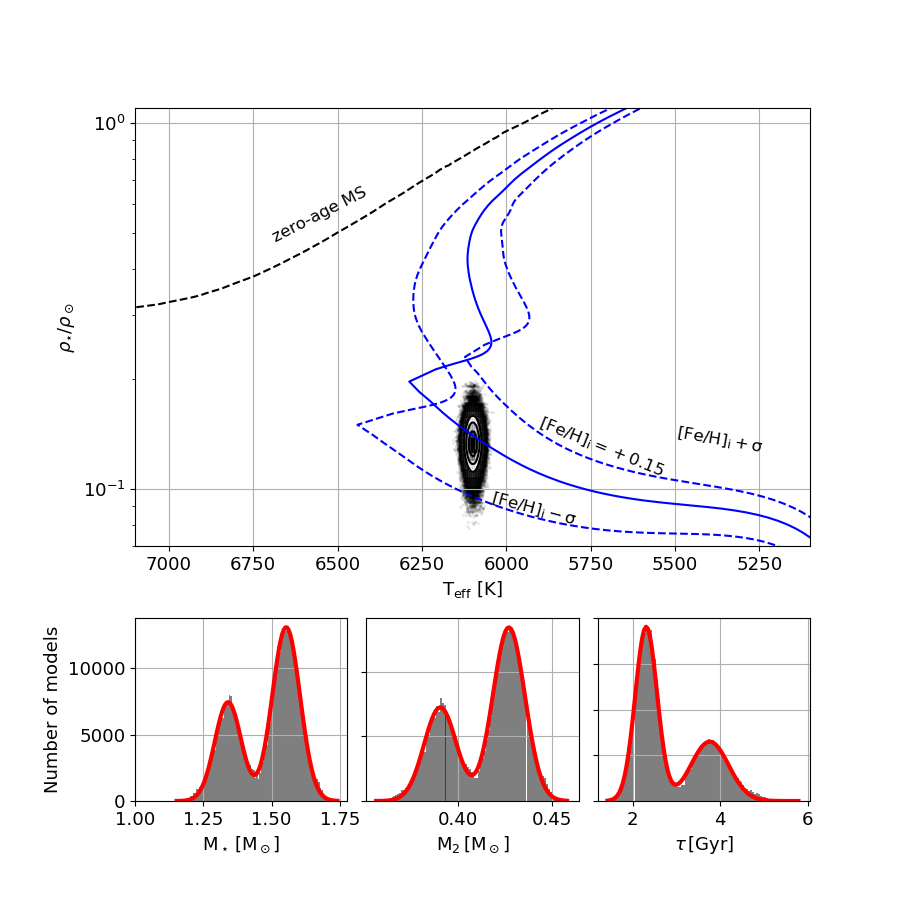}
  \caption{The PPD for the density and temperature of the primary star in J0218$-$31 is shown in the top panel. The zero-age main sequence is show(black-dashed) along with the best fitting isochrone (blue-solid) and the respective isochrones for $\pm 1$-$\sigma$ in [Fe/H]. The lower panels show the PPD distributions for $M_1$, $M_2$ and $\tau$ with best-fitting double-Gaussian models in red.}
  \label{WASP0218-31_HR}
\end{figure*}

J0218$-$31 was observed over three years by the WASP survey. We found a tentative detection of spot-induced variation across the three seasons ($P_{\rm rot}=2.30$\,d, $2.14$\,d and $2.60$\,d). Each have a false alarm probability $<10^{-5}$ and an amplitude around $1$\,mmag amongst a complex periodogram of similar (but smaller) amplitudes making it unclear whether this is due to spot-induced variations ($P_{rot} = 2.35 \pm 0.20$\,d) or poor-quality photometry. From the WASP photometry, we estimated $\Delta t_{\rm tr}= 0.241$\,d and $\Delta m = 0.03$\,magnitudes, corresponding to $R_{1}/a  \approx 0.09$ and $k \approx 0.18$.

We obtained a good SED fit ($\chi_{\rm red}^2 = 0.75$) with the effective surface temperature of the primary star consistent with a spectral type F9 ($T_{\rm eff} \approx 6020$\,K). J0218$-$31 is included in the second data release of Gaia ($G = 9.734$; Source ID 4971670729566470528) with a parallax measurement of $3.762 \pm 0.092$\,mas $ (265.82 \pm  6.50$\,pc). There are 3 close and faint companions within 22" at position angles 266$^\circ$, 332$^\circ$ and 92$^\circ$. The brightest at position angle 332$^\circ$ has $G = 17.140$ ($\Delta G = 7.406$) which would have a negligible flux contribution to the aperture of the WASP photometry and the follow-up R-band photometry. The neighbours at positional angles 266$^\circ$ ($G = 18.128$) and 92$^\circ$ ($G = 19.813$) are fainter still. A brighter companion ($G = 16.015$; source ID 4971670935725243904) is located 50" away at a position angle of 330$^\circ$. This does not overlap the sky annulus of the 1-m SAAO photometry and will have a negligible flux contribution to the WASP photometry. The proper motions of these stars are not similar to J0218$-$31 and so we concluded that they are not physically associated.

We co-added fifty-three out-of-transit spectra to produce a spectrum with S/N $=30$. Using wavelet decomposition, we estimated $T_{\rm eff} = 6100 \pm 85$\,K confirming a spectral class of F9 from SED fitting. The effective temperature is 1-$\sigma$ hotter than predicted by SED fitting suggesting there could be some additional reddening that is unaccounted for. The iron content is higher than the Sun ([Fe/H] = $0.15 \pm 0.12$\,dex). There is also a strong Li\,I  line in the spectrum from which we measured $\log A_{\rm Li}+12= 3.24 \pm 0.08$ suggesting that the convective shell of J0218$-$31 may be similar to that of J2349$-$32.

We fitted the Rossiter-McLaughlin measurements alongside the out-of-transit radial velocity measurements with $g'$, $r'$, $i$ \& $z'$ band photometry to obtain the best fitting orbital solution ($\chi^2_{\rm red} = 1.68$; Fig. \ref{WASP0218-31-orbit}). We initially fitted an independent value of $k$ to the $g'$, $r'$, $i$ \& $z'$ follow-up photometery. The fitted value of $k$ for each bandpass agreed with each other to 1-$\sigma$ suggesting there is no wavelength-dependent transit depths which may have indicated a source of third light. However, we do find a significant drift in systematic velocity ($d(\gamma)/dt = -69.9 \pm 4.1$\,m\,s$^{-1}$yr$^{-1}$) which suggests there may be a faint third body in the system. With the addition of R-M measurements, we were able to calculate the sky-projected angle between the rotational and orbital axes, $\lambda = 4 \pm 7 ^\circ$, which is consistent with the assumption that these axes are aligned. From this we also measured $V \sin i =  10.28 \pm 2.12\, \rm km\,s^{-1}$ which is in agreement with the value inferred from wavelet decomposition.

Similarly to J2308$-$46, J0218$-$31 has entered the ``blue-hook'' part of it's post main-sequence evolution resulting in double-peaked PPDs of $\tau$, $M_{\star}$ and $M_2$. We used the same approach for J2308$-$46 to fit a double-Gaussian to the PPDs for $\tau$, $M_{\star}$ and $M_2$ and found that the younger solution ($2.4 \pm 0.25$ Gyr, $M_{\star} = 1.55 \pm 0.05\,M_{\odot}$, $R_{\star} =  2.13 \pm 0.09\,R_{\odot}$) is favoured with almost twice the likelihood $\mathcal{L}(2.35\, \rm Gyr) / \mathcal{L}(3.80\, \rm Gyr)$ $\approx 3.55$ of the older solution. This is moderate evidence to suggest the younger solution is favoured but we report both solutions in Table \ref{results_table} as a precaution. 


\subsection{EBLM J1847$+$39}

\begin{figure}[htb]
  \centering
  \includegraphics[width=0.9\textwidth/2]{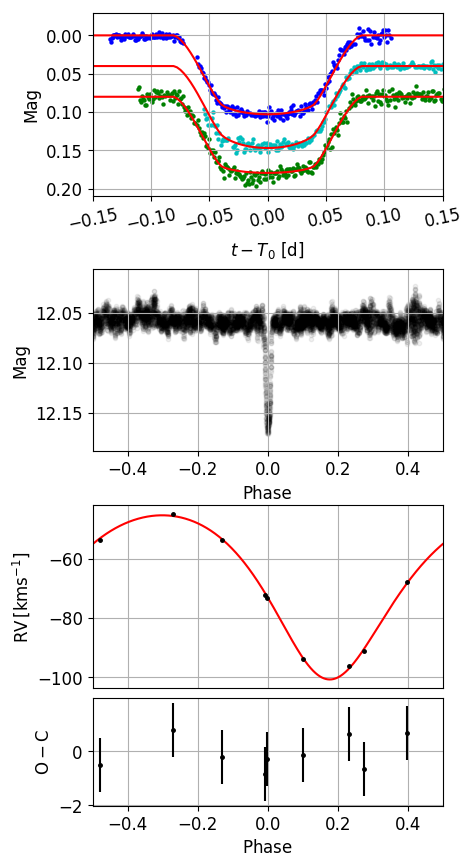}
  \caption{Orbital fit of J1847$+$39. (top panel) Single transits from the HAO in filters CBB (blue), $g'$ (cyan) and $z'$ (green) with best fitting models (red). (upper-middle panel) The phase-folded WASP lightcurve. (lower-middle panel) Drift-corrected radial velocity measurements (black) with the best fitting model (red) and residuals (lower panel).}
  \label{WASP1847+39-orbit}
\end{figure}

J1847$+$39 was observed for three years with the WASP survey. For each season, we found significant spot-induced variations at periods $7.55$\,d, $7.14$\,d and $7.17$\,d with amplitudes of  3\,--\,4\,mmag; each of these signals has a false alarm probability $<10^{-5}$ and we assumed this to be a detection of rotational spot modulation at a period of $7.29\pm0.19$\,d. We find no evidence for ellipsoidal variation in the WASP lightcurve from which we estimated $\Delta t_{\rm tr} = 0.02$\,d and $\Delta m = 0.10$ mag corresponding to initial estimates of $R_{\star}/a \approx 0.07$ and $k\approx0.32$.

The best SED fit estimated the primary star to be of spectral type F9 with temperature of $6020 \pm 100$\,K ($\chi_{\rm red}^2 = 1.37$). This system was included in Gaia DR2 ($G = 11.677$; source ID 2098283457595740288) with a parallax of $3.583 \pm 0.086$\,mas ($279.10 \pm 6.70$\,pc). The field surrounding J1847$+$39 is relatively more crowded compared to the other targets, with over 7 targets brighter than $G=17$ within 1.5'. The closest companion is 12" away ($G=17.694$; source ID 2098283457595821440) at a position angle of $196^\circ$. This neighbour would not have been included in the aperture of J1847$+$39 (radius $\sim$ 6"). A magnitude difference of $\Delta G = 6.17$ results in less than 0.3\% flux contribution if it was included in the apertures of the HAO photometry. There are two bright companions 1.28' and 1.07' away at position angles of $48^\circ$ ($G = 11.577$) and $54^\circ$ ($G=12.135$) respectively. These are beyond the sky annulus of WASP but may still contribute a small amount to the total flux. The proper motions of these objects are dissimilar to J1847$+$39 and so we concluded that they are not associated.

Ten INT spectra were co-added to produce a spectrum with S/N$ = 30$. We used the spectral synthesis method on the wings of the  H$\alpha$ line to estimate $T_{\rm eff} = 6200 \pm 100$\,K (spectral type F8) which is consistent with the SED fit. We were able to fit 11 un-blended Fe\,I lines in the region around H$\alpha$ from which we measured [Fe/H] for each line. We took the mean of value of [Fe/H] as the iron abundance measurement with the standard deviation as the uncertainty ([Fe/H] = $-0.25 \pm 0.21\, \rm dex$). We were unable to determine $\log g$ due to the limited wavelength coverage of the H1800V grating so we assumed $\log g = 4.44$ for the aforementioned synthesis and interpolation of limb-darkening coefficients.

Radial velocity measurements were fitted simultaneously with single transits in $CBB$, $g^{'}$ and $z^{'}$ filters to obtain the best fitting orbital solution ($\chi_{\rm red}^2 = 1.77$; Fig. \ref{WASP1847+39-orbit}). J1847$+$39 has the most eccentric orbit of the sample ($e = 0.209 \pm 0.014$). We attempted to fit an independent value of $k$ for photometry in each filter and found them all to agree within $1$-$\sigma$ suggesting there is no significant third-light contamination. However, we do measure $d(\gamma)/dt = -71.9 \pm 21.7$\,km\,s$^{-1}$yr$^{-1}$ suggesting that there may be a faint third-body in the system. The best-fitting limb-darkening temperature, $T_{\rm eff,ld}$, is $\sim 600$\,K hotter than spectroscopic and photometric analysis; the reason for this is unclear.

The best fitting solution from \textsc{eblmmass} describes a star similar to the Sun in mass and size, but a fifth of it's age ($\tau = 1.10 \pm 1.80$\,Gyr). The systems eccentricity may be primordial in origin as there would have been insufficient time for tidal interaction to circularise the orbit.  The M-dwarf's mass is in the convective transition ($\sim 0.35\,M_\odot$) and provides an interesting test for low-mass stellar models in a region that is highly debated. 


\subsection{EBLM J1436$-$13}

\begin{figure}[htb]
  \centering
  \includegraphics[width=1.\textwidth/2]{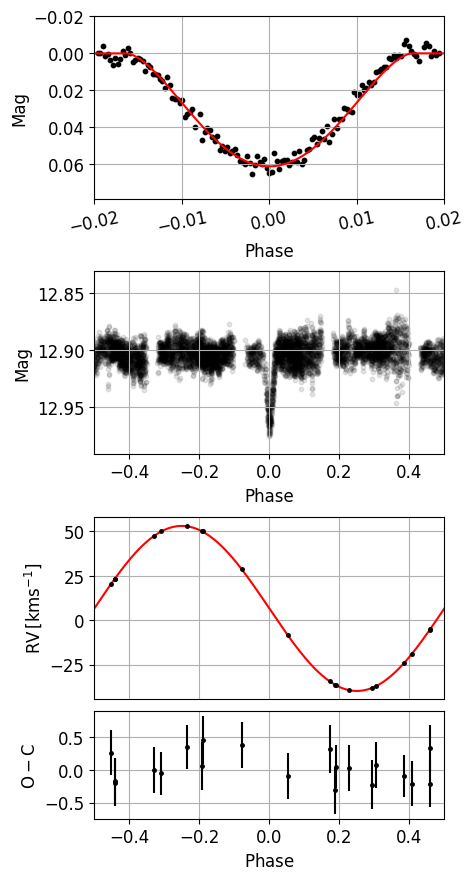}
  \caption{Orbital fit of J1436$-$13. (top panel) A single transit obtained from SAAO in R filter (black) and the best fitting transit model (red). (upper-middle panel) The phase-folded WASP lightcurve. (lower-middle panel) Drift-corrected radial velocity measurements (black) with best fitting model (red) along with residuals (lower panel).}
  \label{WASP1436-13_orbit}
\end{figure}

J1436$-$13 was observed over 3 consecutive seasons with the WASP survey. We found significant variability for each season at periods of $3.99$\,\rm d, $3.98$\,\rm d and $4.02$\,\rm d with amplitudes between 4\,--5\,$\rm mmag$. Each have a false alarm probability $<10^{-5}$ and we assumed this is a detection of spot modulation corresponding to a rotational period $P_{\rm rot} = 4.00\pm 0.02$\,d. We found no evidence for ellipsoidal variation in the WASP photometry from which we measured $\Delta t_{\rm tr} = 0.188$\,d and $\Delta m = 0.065$\,magnitudes corresponding to initial estimates of $R_{\star}/a \approx 0.148$ and $k \approx 0.256$.

The SED fitting measured the primary star to have a spectral type of F9 ($T_{\rm eff} = 6080 \pm 355\,K$; $\chi^2_{\rm red} = 0.87$) with little reddining ($E(B-V) \leq 0.055$ to $1$-$\sigma$). This system is included in Gaia DR2 ($G=12.334$; source I.D 6323183619200685824) with a parallax of $2.063 \pm 0.097$\,mas ($484.73 \pm 22.79$\,pc). There is a faint ($G=18.994$) background star included in Gaia DR2 that is 17" away at a position angle of 302$^\circ$. This is included in the WASP aperture but contributes less than 0.1\% of the total flux.

Thirteen CORALIE spectra were co-added to produce a spectrum with S/N=$30$. Wavelet decomposition measured a value of $T_{\rm eff}$ that is around $300$\,K hotter than predicted by SED fitting suggesting that there may be some unaccounted reddening. The iron content is slightly less than the Sun ([Fe/H]$ = -0.10 \pm 0.12\, \rm dex$) and the magnesium lines are relatively narrow suggesting a low surface gravity. We were unable to identify any measurable lithium lines.

The best fitting orbital solution ($\chi_{\rm red}^2 = 1.75$) describes a transit with a high impact parameter ($b = 0.86 \pm 0.07$; Fig. \ref{WASP1436-13_orbit}). The limb-darkening temperature agrees better with SED fitting than wavelet decomposition, but is consistent with both to 1-$\sigma$. Radial velocity measurements suggest the system is circularised ($e \leq 0.004$ to $1$-$\sigma$) and there is no significant drift in systematic velocity. J1436$-$13 is slightly larger and more massive than the Sun. The uncertainty in $R_\star$ (5\%) is the largest in the sample owing to poorly constrained values of $R_\star / a$ and $k$ owing to a high impact parameter. The M-dwarf companion is the most massive of the sample ($M_2 = 0.49\,M_\odot$).


\section{The mass-radius diagram}\label{discuss:mass-radius}

\begin{figure*}[htb]
  \centering
  \includegraphics[width=1.\textwidth]{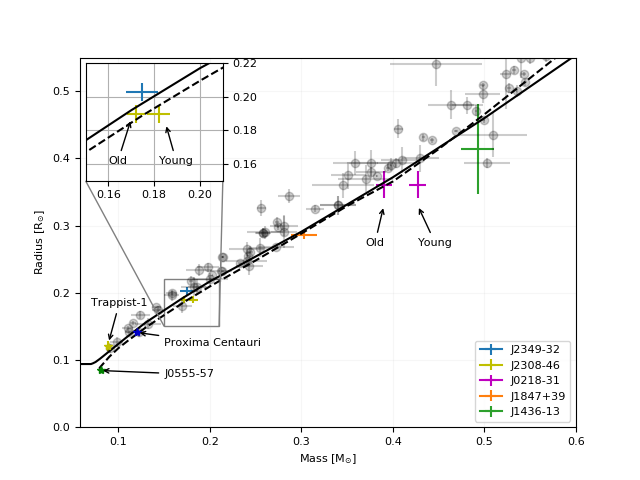}
  \caption{The masses and radii of M-dwarfs in five EBLM systems. The $5\,\rm Gyr$ isochrone for solar metallicity from \protect\citet{2015A&A...577A..42B} is plotted (black-solid) along with the  $5\,\rm Gyr$ isochrone for [M/H]$=-0.5$ \protect \citep{1998A&A...337..403B} (black-dashed). We plot low-mass M-dwarfs with masses and radii known to better than 10\% \protect\citep[from Table 4 of ][ and references therein]{2018arXiv180505841C}. For J2308$-$46 and J0218$-$31 we plot both solutions and label accordingly. We also plot TRAPPIST-1 \protect\citep{2018MNRAS.475.3577D}, Proxima Centauri \protect\citep{Anglada-Escude2016} and J0555$-$57 \protect\citep{vonBoetticher2017}. }
  \label{mass-radius-plot}
\end{figure*}


In Fig. \ref{mass-radius-plot}, we plot the 5\,Gyr isochrones for [Fe/H]$=0$\,dex (B15; \citealt{2015A&A...577A..42B}) and [Fe/H]$=-0.5$\,dex (B98; \citealt{1998A&A...337..403B}) and compare them to the five EBLMs measured in this work. The B15 isochrones rectify some of the flaws in the models presented by \citet{1998A&A...337..403B} (e.g. optical colours that are too blue). Visual inspection of the radii shows that they are broadly consistent with evolutionary models. The EBLMs in this work mostly have a sub-solar metalicity and are expected to have radii between or about the B98 and B15 models.

Two EBLMs (J0218$-$31 and J1436$-$13) have high impact parameters leading to a larger uncertainty in $R_\star / a$, $k$ and ultimately $R_1$ and $R_2$. The effect is most significant in J1436$-$13 where the uncertainties in $R_2$ span across both B98 and B15 isochrones. The primary stars of two EBLMs (J2308$-$46, J0218$-$31) have evolved into the ``blue-hook'' part of their post main-sequence evolution, leading to two solutions of $M_\star$, $M_2$ and $\tau$. Although a single solution is marginally favoured for each, both are valid and we report both in Table \ref{results_table} and Fig. \ref{mass-radius-plot} as a precaution.

%
%
\section{Bayesian measurements of radius inflation}\label{discuss:inflation}

The traditional approach of interpolating between solar B98 ([Fe/H] = 0) and B15 ([Fe/H] = -0.5) isochrones of fixed age is not sufficient to assess inflation, especially for young systems below $1\,$Gyr which may still be contracting. A recent and well-sampled set of isochrones for low-mass stars are required to assess if the M-dwarf in each EBLM system is consistent with the isochrone for the respective measurement of [Fe/H] and $\tau$. For this task, we used MESA stellar evolution models. The MESA models are created using the protosolar abundances recommended by \citet{2009ARA&A..47..481A} as the reference scale for all metallicities; this is consistent with the grid of spectra from wavelet analysis \citep{Gill2018}. MESA uses the OPAL equation of state tables from \citet{2002ApJ...576.1064R} along with opacity tables from \citet{2008ApJS..174..504F}, \citet{2011MNRAS.413.1828Y} and \citet{2010MolPh.108.2265F}. MESA also includes complex treatments for microscopic diffusion and gravitational settling (both important for low-mass stars), radiative levitation (important for high-mass stars), rotation, convective overshooting, magnetic fields and mass-loss. 

\begin{figure*}
    \centering
    \includegraphics[width=1\textwidth]{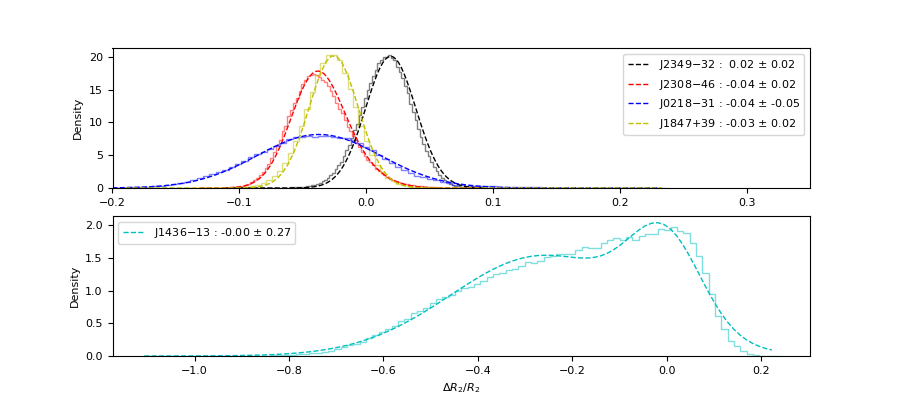}
    \caption{The fractional radius residual PPD for J2349$-$32, J2308$-$46, J0218$-$31 and J1847$+$39 system along with Gaussian models (upper panel). The lower panel shows J1436$-$13 which has a broader PPD and requires a double Gaussian model. The legends denote the best-fitting value of the radius residual with uncertainty. }
    \label{discussion:fig:radius_inflation}
\end{figure*}

The web interpolater \footnote{http://waps.cfa.harvard.edu/MIST/interp\_isos.html} was used to create a grid of MESA isochrones  spanning the range [Fe/H]$=-2$ to $+0.5$\,dex in steps of\,0.5 dex and age range 0.8-9\,Gyrs in steps of 0.2 Gyrs. Using this grid, we created a bi-linear interpolation routine (in dimensions of $\tau$ and [Fe/H]) to obtain an expected radius, $R_{2,\rm exp}$ for a given mass. To assess inflation, the following procedure was employed for each draw in each Markov chain from \textsc{eblmmass} and the orbital solution:

\begin{enumerate}
    \item $\log g_2$ can be calculated from the orbital solution,
    \begin{equation}
    g_2 = \frac{2 \pi}{P} \frac{\sqrt{1-e^2} K_{\star}}{(\frac{R_2}{a})^2 \sin i} .
    \end{equation}
    
    \item The corresponding draw for $M_2$ can be combined with $\log g_2$ to obtain the calculated value of $R_2$,
    \begin{equation}\label{r2_from_m2_and_g2}
        R_2 = \sqrt{\frac{G M_2}{g_2}}.
    \end{equation}

    \item The corresponding draw for $\tau$ was used with a random value for [Fe/H] to interpolate a MESA isochrone. The random value of [Fe/H] was drawn from a Gaussian distribution of mean and width corresponding to the measurement of [Fe/H] and uncertainty of [Fe/H] reported in Table \ref{results_table}.
    
    \item The corresponding draw of $M_2$ was used to interpolate an expected radius for the M-dwarf companion, $R_{2, \rm exp}$. 
    
    \item $R_{2, \rm exp}$ and $R_2$ can be combined to calculate the fractional radius residual,
    \begin{equation}
        \frac{\Delta R_2}{R_2} = \frac{R_2 - R_{2,exp}}{R_2}.
    \end{equation}
\end{enumerate}

By repeating the above procedure for each draw in each Markov chain from \textsc{eblmass} and the orbital fit, we were able to estimate the PPD for the fractional radius residual for each EBLM (Fig. \ref{discussion:fig:radius_inflation}). Four EBLMs have narrow-peaked PPDs for $\Delta R_2 / R_2$ (top panel of Fig. \ref{discussion:fig:radius_inflation}). For these, we calculated the nominal fractional radius by binning the PPD into 100 bins and fitted a Gaussian model; we took the mean of the fitted Gaussian to be the measurement of $\Delta R_2 / R_2$ with uncertainty equal to the standard deviation. We found that a Gaussian shape is not a perfect fit to the PPDs of $\Delta R_2 / R_2$; there are asymmetric discrepancies where one side of the Gaussian model is lower than the PPD, whilst the other is too high. On average, the under-prediction on one side and over prediction on the other are of the same magnitude and we assume the widths still accurately represent the mean uncertainty of $\Delta R_2 / R_2$. 

J1436$-$13 has a significantly higher impact parameter which broadens the PPD for $R_2$ and thus, $\Delta R_2 / R_2$ (lower panel of Fig. \ref{discussion:fig:radius_inflation}). We approximate this shape with a double Gaussian, and use an identical routine used to measure the double-peaked PPDs in Sect. \ref{results:J2308}. The fit for J1436$-$13 is not perfect and we used the peak of the PPD as the measurement of $\Delta R_2 / R_2$ with uncertainty equal to the standard deviations of each fitted Gaussian added in quadrature. 

J2308$-$46 and J1847$+$39 appear deflated by at least 1-$\sigma$ compared to MESA evolutionary models. There is also some evidence to suggest J0218$-$31 is deflated although the measured radius is consistent with predictions from evolutionary models to 1-$\sigma$. Conversely, J2349$-$32 appears inflated by 1-$\sigma$. The PPD for $\Delta R_2 / R_2$ for J1436$-$13 is broadly consistent with evolutionary models although it is not possible to assess inflation for this since the radius is so poorly constrained. J2308$-$46 and J0218$-$31 have double-peaked distributions for $M_2$ and $\tau$ and we expected the PPDs for $\Delta R_2 / R_2$ to be shaped similar since $M_2$ is used to calculate $R_2$, and combined with $\tau$ to estimate $R_{\rm 2,exp}$. In creating the PPD for $R_2$ (Eqn \ref{r2_from_m2_and_g2}), the division of the PPD for $M_2$ with the PPD for $g_2$ diminishes the double-peaked nature observed in the PPD $M_2$, leading to a Cauchy-like PPD for $R_2$. The interpolated value of $R_{2, \rm exp}$ is dependent on $\tau$ and $M_2$ which are both double peaked. $R_{2, \rm exp}$ is not expected to have a double-peaked PPD as each combination of $\tau$ and $M_2$ was a trial step in \textsc{eblmmass} and will correspond to a similar expected radii (i.e. higher values of $\tau$ will correspond to lower values of $M_2$ and vice-versa). Thus the PPD for $\Delta R_2 / R_2$ is single peaked with width controlled by the uncertainty in $M_2$, $g_2$ and [Fe/H].

\section{Systematic effects  on determining mass, radius and age}\label{discussion:systematics}

One major issue remains with the method employed in this paper and previous publications of the EBLM project: we are attempting to test evolutionary models of low-mass stars using the models of better-understood F-dwarfs. This method is acceptable when the uncertainty propagated by stellar models for FG stars are much smaller than the uncertainties of physical properties propagated by observational uncertainties in radial velocity and transit photometry measurements. We must also consider the possibility of unresolved, faint components in the aperture which could systematically modify the transit shape along with physical properties inferred from modelling. 
In the following sections we explore the scale of some potential sources of additional uncertainties on measurements of physical properties for EBLM systems. For J2308$-$46 and J0218$-$31, we only considered the most probable solution in this section. 

\begin{sidewaystable*}
\caption{The difference in mass and radius of the primary star ($\star$) and the secondary (2) for a variety of different scenarios. The measured values were subtracted from those in Table. \protect{\ref{results_table}}). For J2308-46 and J0218-31, we only considered the most probable solution in this work. We separately re-fitted with 10\, \% third light ($l3$) and using the quadratic limb-darkening law over the Claret law ($ldy$) from which we re-measured only the radii of the stars in the systems. We also separately recalculate the masses of the stars in each system by changing mixing length parameter from $1.50$ to $1.78$ ($\alpha_{mlt}$) and a change in helium enhancement values from $0.00$ to $0.02$ ($\Delta Y$). We also show the mean of each column, $\bar{x}$, calculated with all values excluding those from J1436-13 (marked with an asterisk).}              
\label{systematic_light_table}      
\centering                                    
\begin{tabular}{c l l  l l l l l l l l l l l l l }          
\hline\hline 
 & $\Delta R_{\star,\, l3}$ & 
 $\Delta R_{2,\, l3}$ 
 &$\Delta R_{\star,\, ldy}$ & 
 $\Delta R_{2,\, ldy}$ 
 
 & $\Delta M_{\star,\, \alpha_{mlt}}$
 & $\Delta M_{2,\, \alpha_{mlt}}$
 & $\Delta \tau_{\alpha_{mlt}}$

 & $\Delta M_{\star,\, Y}$
 & $\Delta M_{2,\, Y}$
 & $\Delta \tau_{Y}$

  \\
\hline

J2349-32 & $-0.003$ & $0.016$ & $-0.014$ & $-0.002$ & $-0.037$ & $-0.004$ & $2.208$ & $0.048$ & $0.005$ & $-0.853$ & \\
& ($-$0.22\%) & (7.8\%) & ($-$1.46\%) & ($-$0.98\%) &  ($-$3.73\%) & ($-$2.30\%) & (96\%) & (4.84\%) & (2.87\%) & ($-$37\%) \\

J2308-46 
& $-0.002$ 
& $0.013$ 
& $-0.014$ 
& $-0.001$ 
& $-0.044$ 
& $-0.002$ 
& $1.423$ 
& $0.059$ 
& $0.005$ 
& $-1.32$ & \\
& ($-$0.13\%) 
& (3.60\%)
& ($-$0.91\%)
& ($-$0.28\%)
& ($-$3.59\%)
& ($-$0.46\%)
& (59\%)
& (4.82\%)
& (1.17\%) 
& ($-$55\%) \\

J0218-31 & $-0.007$ & $0.012$ & $-0.005$ & $-0.002$ & $-0.057$ & $-0.001$ & $0.503$ & $0.041$ & $0.006$ & $-0.254$ \\

& ($-$0.32\%)& (3.32\%)& ($-$0.23\%)& ($-$0.55\%)& ($-$3.68\%)& ($-$0.23\%)& (13\%)& (2.65\%)& (1.41\%)& ($-$7\%) \\

J1847+39 & $-0.008$ & $0.016$ & $-0.015$ & $-0.004$ & $-0.046$ & $-0.001$ & $1.071$ & $0.052$ & $0.009$ &  $-0.820$ \\

& ($-$0.90\%)& (5.57\%)& ($-$1.50\%)& ($-$1.39\%)& ($-$4.36\%)& ($-$0.33\%)& (97\%)& (4.93\%)& (2.97\%) & ($-$74\%) \\
	
J1436-13* & $-0.162$ & $0.064$ & $-0.059$ & $-0.090$ & $-0.024$ & $-0.005$ & $-0.135$ & $0.043$ & $0.011$ & $-1.150$ \\

& ($-$11.91\%) & (15.68\%)& ($-$4.34\%)& ($-$22.05\%) & ($-$2.03\%) & ($-$1.02\%)& ($-$6\%) & (3.62\%) & ($-$2.24\%) & ($-$50\%) \\

\hline 
$\bar{x}$ & $-$0.39\% & 5.07\% & $-$1.02 \% & $-$0.80 \% & $-$3.84 \% &  $-$0.83 \% & 63\%  & 4.31\% & 2.11\% & 43\% \\

\hline
\end{tabular}
\end{sidewaystable*}

\subsection{Evolution ambiguity, $\alpha_{\rm MLT}$ and $Y_{\rm He}$}
The default model grid used in \textsc{eblmmass} uses a mixing length parameter $\alpha_{MLT} = 1.78$ and an initial helium abundance $\rm Y = 0.26646 + 0.984 \, \rm Z$, both of which have been calibrated on the Sun. As noted by \cite{2015A&A...575A..36M}, these assumptions are subject to some level of uncertainty.  \citet{Maxted2015d} estimated the additional uncertainty in $M_{\star}$ and $\tau$ for 28 transiting exoplanet host stars by assuming an error of 0.2 in $\alpha_{MLT}$ and 0.02 for $\Delta Y$.   They find that systematic errors in $M_{\star}$ and $\tau$ from $Y$ and $\alpha_{MLT}$ can be comparable to the random errors in these values for typical observational uncertainties in the input parameters. We note that the sample used in \cite{Maxted2015d} consists primarily of stars less massive than the Sun, whereas the primary stars in this work are more massive F-type stars.  Three grids of models are provided with \textsc{eblmmass}: 1. $\alpha_{MLT} = 1.78$, $\Delta Y=0.00$, 2. $\alpha_{MLT} = 1.5$, $\Delta Y=0.00$ and 3. $\alpha_{MLT} = 1.78$, $\Delta Y=0.02$; we used grid 1 in Table \ref{results_table}. We re-measured the mass, radius and age of both components with the grids 2 and 3 to see how the uncertainties in $\alpha_{\rm MLT}$ and $\Delta Y$ impact our results. We used grid 2 to assess an additional uncertainty of 0.28 for $\alpha_{MLT}$ and grid 3 to assess an additional uncertainty of 0.02 for $\Delta Y$. We used the same orbital solution and atmospheric parameters reported in Sect. \ref{results} and report our results in Table \ref{systematic_light_table}. 

We found that an additional uncertainty in $\alpha_{\rm MLT}$ results in up to a 4\% increase in the mass uncertainty. This is mirrored by the 1-5\% increase in the mass uncertainty seen for an additional uncertainty of 0.02 for $\Delta Y$. These results are consistent with those found by \cite{Maxted2015d} and the largest increase in uncertainties are seen for the host stars. The measured age of each system has a larger fractional error than the measured masses and we can expect and additional uncertainty if $\alpha_{\rm MLT}$ and $\Delta Y$ are unconstrained. Assuming that both parameters are poorly constrained, we can expect an additional 3-5\% uncertainty in the mass of the M-dwarf companion. This is significant and has potential to skew the interpretations of the mass-radius diagram.

A further limitation arises when the primary star has evolved into the post-main sequence blue hook (Henyey hook). J2308$-$46 and J0218$-$31 are in this region leading to two distinct solutions for $M_{\star}$ and $\tau$. A single solution is preferred for both these systems but there will always be some ambiguity until further mass constraints can be obtained. A solution may lie in the increased contrast between a FGK star and an M-dwarf in the infrared. It may be possible to detect molecular lines (VO, TiO and CaH) associated with an M-dwarf using high-resolution infrared spectroscopy. Simultaneously cross-correlating templates from G-/M-dwarfs with near-infrared spectra of EBLMs should produce similar results to what \citet{2012ApJ...751L..31B} achieved in the optical for Kepler-16. The act of turning an SB1 into and SB2 would provide a direct
test of the methods used in this work. This would also place a further constraint on which mass and age solution best describes the system in cases whereby the primary star is close to the post main-sequence blue-hook. Inspection of where these targets sit in the colour-magnitude diagram (Fig. \ref{methods:fig:gaia_EBLMsmy_label}) reveal that J2308$-$46 and J0218$-$31 are significantly closer to the giant branch than the other three EBLMs. An arbitrary cut using Gaia colours could be used to pre-select cooler host stars ($\leq 6100$ K) and avoid host stars near the post-main sequence blue hook. 


\subsection{Third light effect}\label{third_light_effect}

\begin{table*}
\caption{Distance measurements from Gaia DR2. We also report the orbital separation corresponding to an sky-projected separation of 0.3" for each EBLM system and the orbital period associated with this separation using $M_{\star}$ from Table \ref{results_table}.}              
\label{EBLM_seperation}      
\centering                                      
\begin{tabular}{c c c c c}          
\hline\hline                        
EBLM & Parallax [$mas$] & d [$pc$] & Orbital separation at 0.3" [$au$] & Period  [$yr$] \\    
\hline                                   
    J2349-32 & $3.769 \pm 0.092$ & $265.32 \pm 6.47$   & $394 \pm 12$    &  $19.97 \pm 0.27$ \\      
    J2308-46 & $2.187 \pm 0.113$ & $457.24 \pm 23.63$  & $658 \pm 27$   &  $22.07 \pm 0.65$ \\
    J0218-31 & $3.762 \pm 0.092$ & $265.82 \pm 6.50$   & $387 \pm 7$    &  $15.24 \pm 0.10$ \\
    J1847+39 & $3.583 \pm 0.086$ & $279.10 \pm 6.70$   & $421 \pm 9$    &  $19.34 \pm 0.15$ \\
    J1436-13 & $2.063 \pm 0.097$ & $484.73 \pm 22.79$  & $691 \pm 24$   &  $23.64 \pm 0.32$ \\

\hline                                             
\end{tabular}
\end{table*}

Lucky imaging provides constraints on nearby contaminating objects. For J2349$-$32 and J2308$-$46, we find close companions which do not significantly contaminate follow-up photometry. For J0218$-$31 and J1847$+$39 we can put constraints on the amount of third light from the consistency between the ratio of the radii measured from transit photometry in different pass bands. For J1436$-$13 we have to rely on existing surveys to identify any nearby stars which may contaminate follow-up photometry.  Inspection of the Gaia survey DR2 (resolution of $\leq 1\arcsec$) finds no evidence of blends or contamination. Ground-based lucky imaging has a limited resolution to resolve companions with a sky-projected separation of $\sim$0.3\arcsec. \footnote{Determined from the limiting resolution of lucky imaging observations in this work.} The orbital separation for each EBLM corresponding to a sky-projected separation of 0.3\arcsec was calculated using parallax measurements from Gaia DR2 (Table \ref{EBLM_seperation}). The period of such orbits were also calculated using measurements of $M_{\star}$ from Table \ref{results_table}. We find that the closest EBLM (J2349$-$32 at a distance of $259 \pm 3\, pc$) would require a semi-major axis of at least $389\, au$ with orbital period spanning decades. The three-body systems identified by  \citet{Triaud2017} will have orbital periods similar on the order of decades and would be difficult or impossible to resolve through lucky imaging. 

The spectrum itself can provide useful insights for potential aperture contamination. The analysis of CORALIE spectra for 118 EBLM systems presented by \citep{Triaud2017} found that 17.8\,\% of these systems show significant evident for non-zero values of $d(\gamma)/dt$ (spanning $d(\gamma)/dt = 0.07$ $-$4.5 $\rm km\, \rm s^{-1}\, \rm yr^{-1}$). J0218$-$31 and J1847$+$39 have best-fitting values of $d(\gamma)/dt$ which are at the bottom of this bracket.  If these drifts are evidence of a third body, they would have separations which are challenging to resolve with lucky imaging. If they could be resolved with lucky imaging, they would have periods which would require decades of radial velocity monitoring to characterise them (Table \ref{EBLM_seperation}). The longest EBLM system has spectroscopic observations spanning 7.5\,yrs (see Table \ref{Observeration_table}) and so determining the nature of the systematic drift in radial velocities for each system would be difficult. The low signal-to-noise spectra from CORALIE and INT eliminates unresolved blends which contribute more than 30\% of the luminosity of the primary star by inspection of cross-correlation functions (350-750\,nm). 

Including third light as a free parameter in the orbital fit changes the shape and depth of a light-curve leading to degeneracies between $R_{\star}$, $k$ and $b$.  We assessed this by re-fitting the orbital solution for all stars assuming a 10\% light contamination from a third body which does not interact with the EBLM system. From this fit, we combined best fitting values of  $R_{\star}/a$, $b$, and $k$ and their uncertainties with nominal values of the remaining parameters from the original fit to re-determine $R_{\star}$ and $R_{2}$ (first two columns in Table \ref{systematic_light_table}). On average, we find a 3-8\% increase in $R_2$ when third light is fixed to 10\,\%; with the largest uncertainty for the smallest M-dwarfs. We ignore J1436$-$13 from this discussion due to the grazing transit nature. This is comparable to the reported radius inflation for low-mass stars typically quoted in the literature \citep[e.g.  $3-5\%$; ][]{Spada2013}. However, if we were to see radius inflation in general for the M-dwarf components of EBLM systems then the third-light effect can only be a partial explanation. This is because the majority of these systems do not have detected third bodies in the system, and the third body will often contribute much less than 10\% of the total flux in such triple-star systems. 


\subsection{Limb darkening}\label{limb_darkening_section}

\begin{table*}
\caption{Theoretical (marked with an asterisk) and fitted quadratic limb-darkening coefficients for  $a_1$ and $a_2$ using Eqn. \ref{quad_limb_law}.}              
\label{quadratic_limb_darkening__table}      
\centering                                    
\begin{tabular}{c c c c c c}          
\hline\hline 
EBLM   & Filter& $a_1^*$ & $a_1$ & $a_2^*$ & $a_2$ \\
\hline

J2349-32  & I & $0.368 \pm 0.050$  & $0.400 \pm 0.010$ & $0.147 \pm 0.051$ & $0.145 \pm 0.050$    \\

J2308-46 & R & $0.460 \pm 0.050$ & $0.444 \pm 0.031$ & $0.150 \pm 0.051$ & $0.128 \pm 0.043$ \\

J0218-31 & g' & $0.718 \pm 0.051$ & $0.735 \pm 0.022$ & $0.050 \pm 0.052$ & $0.278 \pm 0.013$\\
&          r' & $0.508 \pm 0.050$ & $0.588 \pm 0.012$ & $0.136 \pm 0.052$ & $0.203 \pm 0.015$\\
&          i' & $0.412 \pm 0.050$ & $0.461 \pm 0.011$ & $0.143 \pm 0.051$ & $0.227 \pm 0.014$ \\
&          z' & $0.338 \pm 0.050$ & $0.341 \pm 0.009$ & $0.146 \pm 0.051$ & $0.201 \pm 0.015$\\

J1847+39 & CBB & $0.468 \pm 0.050$ & $0.461 \pm 0.034$ & $0.147 \pm 0.051$ & $0.217 \pm 0.015$ \\
&          g'  & $0.659 \pm 0.051$ & $0.631 \pm 0.057$ & $0.100 \pm 0.051$ & $0.223 \pm 0.015$ \\
&		   z'  & $0.303 \pm 0.050$ & $0.255 \pm 0.035$ & $0.214 \pm 0.050$ & $0.215 \pm 0.022$\\
	
J1436-13 & R & $0.453 \pm 0.050$ & $0.547 \pm 0.010$ & $0.151 \pm 0.051$ & $0.247 \pm 0.015$\\

\hline
\end{tabular}
\end{table*}
To determine accurate estimates for $R_{\star}$, $k$ and $b$ we required an accurate prescription for limb-darkening in our light curve model.  
For this work, we have used the Claret 4-parameter law \citep{Claret2000},
\begin{equation}\label{claret_limb_law}
\frac{I_\mu}{I_0} = 1 - \sum_{i =1}^{4} a_i(1-\mu^{\frac{1}{2}}_i),
\end{equation}
where $a_i$ is the $i^{th}$ limb-darkening coefficient. The coefficient tables we use are provided by \citet{Claret2011} for \textit{Kepler}, Str\"{o}mgren, Johnson-Cousins and Sloan pass bands based on ATLAS stellar atmosphere models assuming a micro-turbulent velocity $\xi = 2 \, \rm km\,s^{-1}$. For the SDSS passbands ($u'$, $g'$, $r'$, $i'$, $z'$), we use the tables from \cite{Claret2004} provided with \textsc{ellc}. We interpolate these tables for a given $T_{\rm eff}$, [Fe/H] and $\log g$ to obtain 4-parameter limb darkening coefficients and a gravity darkening coefficient using the interpolation routine provided with \textsc{ellc}. As described in Sect. \ref{orbital_fit}, we allow the limb-darkening temperature, $T_{\rm eff, ld}$, to vary as a free parameter with a Gaussian prior from spectroscopy, and fix $\log g$ and [Fe/H] to values from spectroscopic analysis. An alternative is to use the quadratic limb-darkening law \citep{Kopal1950} with only 2 parameters,

\begin{equation}\label{quad_limb_law}
\frac{I_u}{I_0} = 1 - \sum_{i =1}^{2} a_i (1 - \mu)^i,
\end{equation}
and allow both coefficients to vary in a fit using the de-correlated parameters $a_+ = a_1 + a_2$ and $a_- = a_1 - a_2$ \citep{Brown2001}. Alternate de-correlation parameters which are not used in this work have been suggested by \cite{Kipping2013}: $q_1 = (a_1 + a_2)^2$ and $q_2 = 0.5 \times a_1 ( a_1 + a_2)^{-1}$.

We assessed the choice of limb-darkening law on $R_{\star}$ and $R_2$ by re-fitting each system using the quadratic limb darkening law (Eqn. \ref{quad_limb_law}). We generate coefficients $a_1$ and $a_2$ for each pass-band using the Python package \textsc{ldtk} \citep[see Table \ref{quadratic_limb_darkening__table}; ][]{Parviainen2015}. \textsc{ldtk} uses uncertainties from $T_{\rm eff}$, [Fe/H] and $\log g$ to estimate uncertainties in the calculated values of $a_1$ and $a_2$ ($\sigma_{a_1}$ and $\sigma_{a_2}$). We use these uncertainties to apply Gaussian priors to $a_1$ and $a_2$ and stop the sampler tending to unrealistic values. These priors have a mean value and uncertainty calculated from \textsc{ldtk}. Errors on $a_1$ and $a_2$ from errors on $T_{\rm eff}$, etc. are very small and unlikely to reflect real uncertainty due to uncertainties in the models so we add a subjective value of 0.05 in quadrature to the uncertainties on each parameter to allow for this. A new combined orbit and light curve solution was found using the same number of draws used in Sect. \ref{results}. From this solution, we use $R_{\star}/a$, $k$ and $b$ with their uncertainties and combine it with nominal parameters from the orbital solution in Sect. \ref{results} to measure the radii of components in each system. This ensures that only parameters relating to the radii of the stars were changed. 

We found that the additional uncertainty introduced by the choice of limb-darkening law (Table \ref{quadratic_limb_darkening__table}) is less than that introduced by third light. The primary and secondary stars see a reduction in $R_{\star}$ and $R_2$ between $0.5-2\%$. \cite{2013A&A...549A...9C} from their study of exoplanet-host stars conclude that fixing the limb-darkening coefficients to theoretical values does not allow the determination of $R_2$ to better than $1$-$10 \%$; a reason why we fitted $a_1$ \& $a_2$. Intertwined in this is the effects caused by stellar activity, spots and faculae. These are time-dependent effects which change at each transit event and can modify the limb-darkening values far from what is predicted. One conclusion from \cite{2013A&A...549A...9C} is that a star with 0.5\% spot coverage can still introduce a 1\% uncertainty on $k$.


\section{Conclusion}

We present the orbital solutions for five F+M binary systems (EBLMs) discovered by the WASP survey. The host stars for these EBLMs are of spectral type F and are predominantly metal deficient with the exception of J0218$-$31. J2308$-$46 has $V\sin i = 39.82 \pm 1.35$\,kms$^{-1}$ and appears to be near the Kraft break which separates stars with deep convective envelopes and efficient dynamos to those without. 

We found variations in the WASP lightcurves of J2349$-$32, J0218$-$31, J1847$+$39 and J1436$-$13 which are similar to the best-fitting orbital period. 
There is a strong $p/2$ signal for J2308$-$46 corresponding to ellipsoidal variation. We fixed parameters associated with ellipsoidal variation ($q=0.05$, gravity darkening coefficient = 0.1) to estimate the starting parameters of $R_\star /a$ and $k$ for the orbital solution. We have showed that assuming a spherical star-shape for J2308$-$46 is a good approximation and does not significantly alter the fitted transit model.

Radial velocity measurements and follow-up photometry were fitted simultaneously to obtain the best-fitting orbital solution (Table \ref{results_table}). All EBLMs except J1847$+$39 ($e = 0.209$) have small eccentricities ($e \leq 0.03$). J0218$-$31 and J1847$+$39 have $d(\gamma)/dt = -69.9 \pm 4.1\,\rm m\,s^{-1}yr^{-1}$ and $-71.9 \pm 21.7 \,\rm m\,s^{-1}yr^{-1}$ which suggests that these systems may be influenced by a faint, unresolved companion. 
Significant trends in the follow-up photometry of J2308$-$46 required a red-noise model to be generated using a Matern-3/2 kernel with fixed values of $\log \rho = 2$ and $\log \sigma = 2$. This produced a more acceptable fit around contact points 1 and 2 than would have been achievable without a red-noise model. J1847$+$39 is moderately eccentric ($e\approx0.2$) and is not circularised.

The best-fitting orbital solution was combined with atmospheric parameters to interpolate evolutionary models (\textsc{eblmmass}) and measure the masses, radii and age of each component in all five systems. The masses of the primary stars span $0.99$-$1.55\, \rm M_\sun$ with radii spanning $0.96$-$2.13\, \rm R_\sun$. The primary stars of J2308$-$46 and J0218$-$31 have evolved into a region near the post-main sequence blue hook resulting in two distinct solutions of masses, radii and age. For both EBLMs, one solution is slightly more favourable than the other but we report both solutions in Table \ref{results_table} as a precaution.
The M-dwarf companions of J2308$-$46 and J1847$+$39 appear deflated by 2-$\sigma$ and 1.5-$\sigma$ respectively. There is moderate evidence to suggest that J0218$-$31 is also deflated despite being consistent with MESA stellar models to within 1-$\sigma$. J2349$-$32 is inflated by 1-$\sigma$. J1436$-$13 has a high impact parameter and it is difficult assess inflation for this target.

We made various choices to measure the masses, radii and ages of both components for each EBLM. One choice was to use the Claret 4-parameter limb-darkening law instead of the  plethora of other laws used in the literature. We re-fit all systems using the quadratic limb-darkening law with theoretical coefficients calculated using \textsc{ldtk}. The majority of stars see a reduction of $R_\star$ and $R_2$ below 2\%. Spectroscopy used in this work can only rule out unresolved companions which contribute >30\% of the total flux. We investigated the effect of 10\% third light when measuring the radii of components in EBLMS. We find that $R_2$ increases by 3-8\% reaffirming the necessity to rule out sources of third-light. The assumptions we make regarding which evolutionary models we use have significant consequences too. Additional uncertainty in $\Delta Y$ can introduce an additional mass uncertainty $\approx 1-5 \%$, while additional uncertainties for $\alpha_{\rm MLT}$ introduce an additional mass uncertainty up to $4 \%$. Assuming that both parameters are poorly constrained, we can expect an additional 3-5\% uncertainty in the mass of the M-dwarf companion.

\bibliographystyle{aa}
\bibliography{references.bib}

\section{Acknowledgements}

SG, PM and JE gratefully acknowledge financial support from the Science and Technology Facilities Council [ST/N504348/1, ST/M001040/1, ST/N504348/1 and ST/P000495/1]. This work has made use of data from the European Space Agency (ESA) mission {\it Gaia} (\url{https://www.cosmos.esa.int/gaia}), processed by the {\it Gaia} Data Processing and Analysis Consortium (DPAC, \url{https://www.cosmos.esa.int/web/gaia/dpac/consortium}). Funding for the DPAC has been provided by national institutions, in particular the institutions participating in the {\it Gaia} Multilateral Agreement.  Results from an initial assessment of the data for these targets conducted by the following undergraduate astrophysics students at Keele University were used as a starting point for the analysis presented here: J. Burrows, M. Hill, C. Huggins, M. Jacobs, T. Le Vien, G.~Phillips, L. Smith, A. Stay, A. Welch. PL was supported by the project ANT1656. This project has received funding from the European Research Council (ERC) under the European Union’s Horizon 2020 research and innovation programme (grant agreement n$^\circ$ 803193/BEBOP). MG is FNRS Senior Research Associate, and acknowledges support from an ARC grant for Concerted Research Actions, financed by the Wallonia-Brussels Federation. This paper uses observations made at the South African Astronomical Observatory (SAAO). We thank the reviewer for his/her thorough review and highly appreciate the comments and suggestions  which significantly contributed to improving the quality of this work.

\begin{appendix}

\section{SED fits}

 \begin{figure*}
\centering
 \begin{subfigure}[b]{0.45\linewidth}
    \centering
    \includegraphics[width=\linewidth]{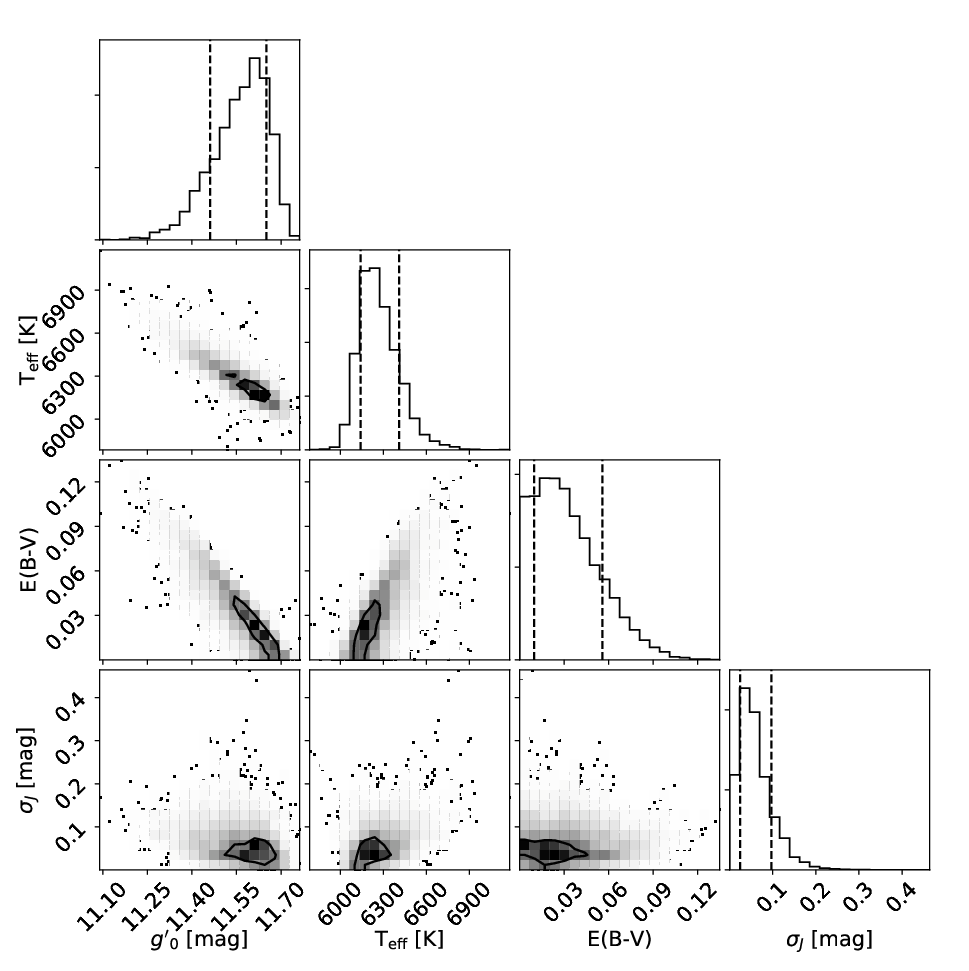} 
    \vspace{4ex}
  \end{subfigure}
  \begin{subfigure}[b]{0.45\linewidth}
    \centering
    \includegraphics[width=\linewidth]{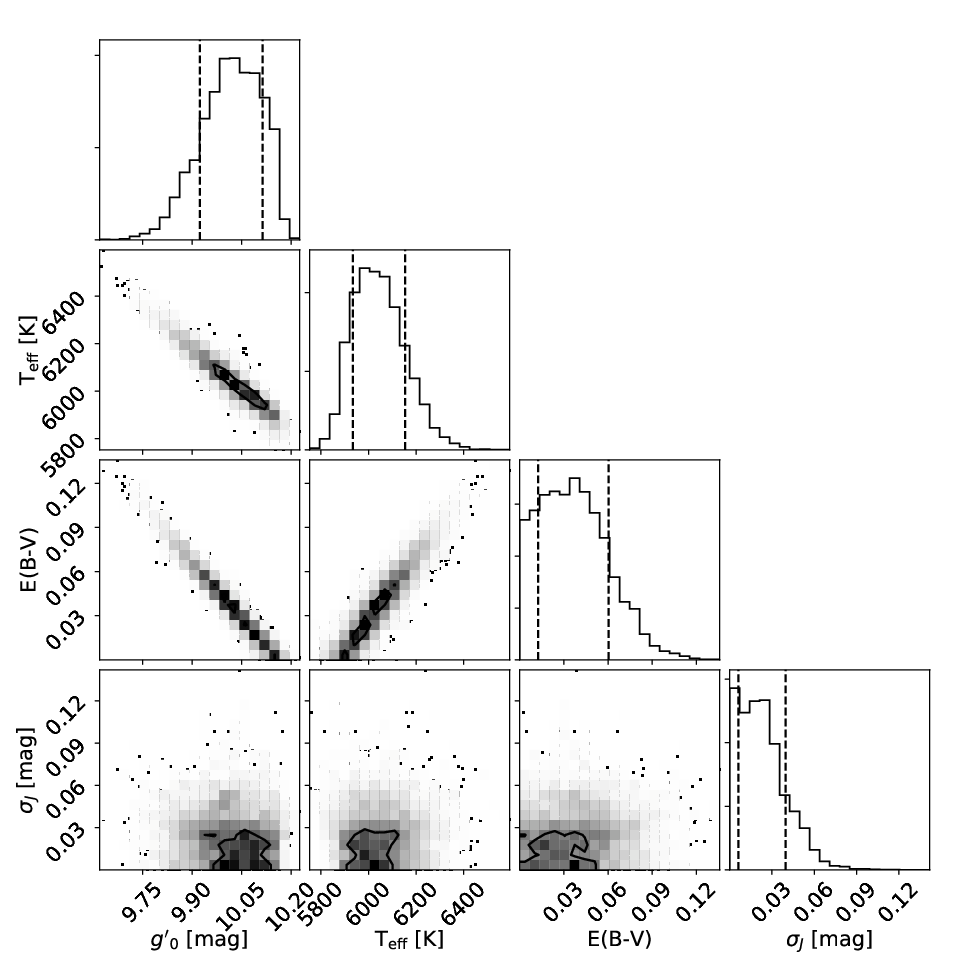}
    \vspace{4ex}
  \end{subfigure} 
  \begin{subfigure}[b]{0.45\linewidth}
    \centering
    \includegraphics[width=\linewidth]{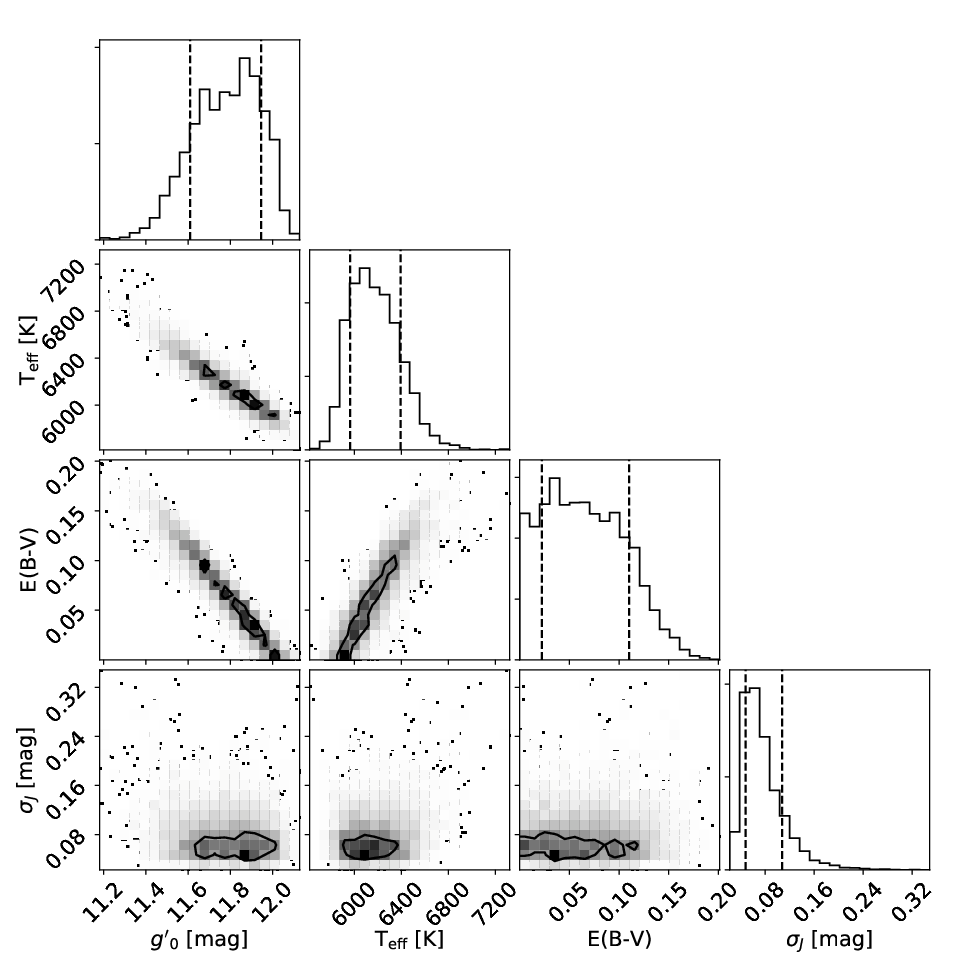} 
  \end{subfigure}
  \begin{subfigure}[b]{0.45\linewidth}
    \centering
    \includegraphics[width=\linewidth]{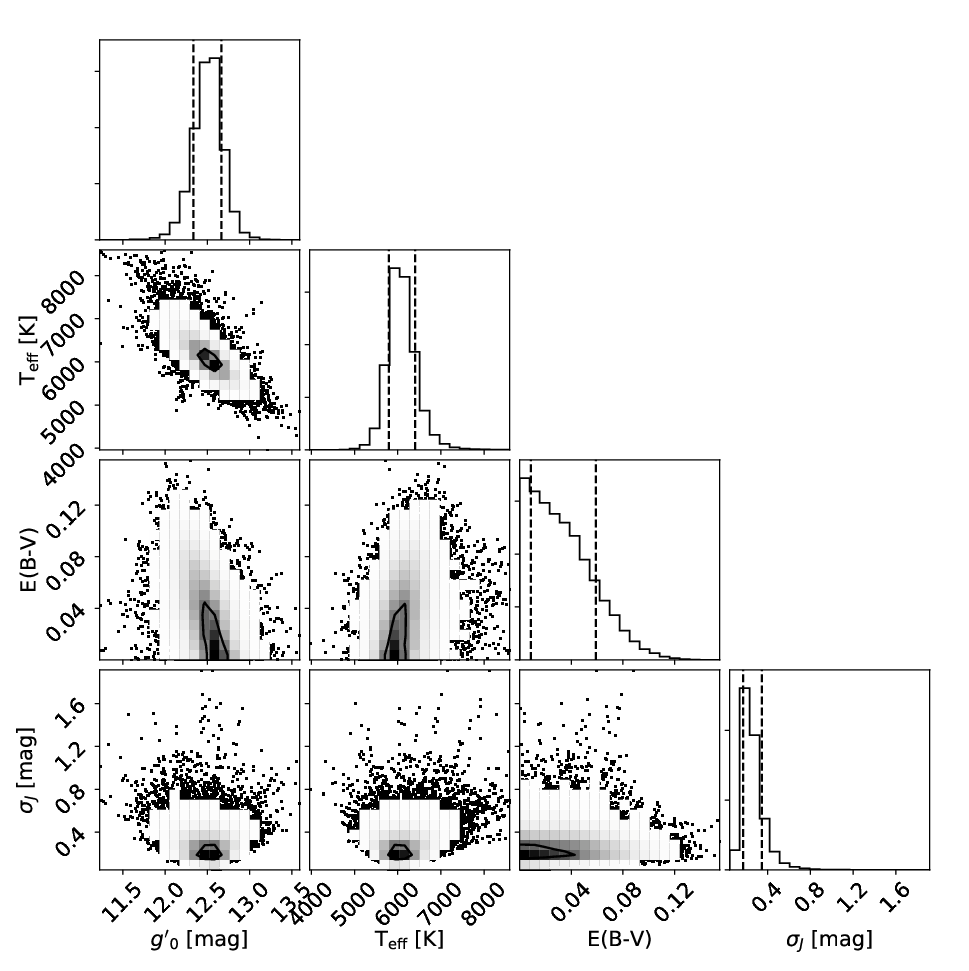}
  \end{subfigure} 
  \caption{The posterior probability distribution of EBLM J2308$-$46 (top-left), J0218$-$31 (top-right), J1847$+$39 (bottom-left) and J1439$-$13 (bottom-right) from photometric fitting. Over plotted are the 1-$\sigma$ contours.}
  \label{SED}
  \end{figure*}

\section{Lomb-Scargle diagrams}\label{appendix:lomb}

\begin{figure*}
    \centering
    \includegraphics[width=\linewidth]{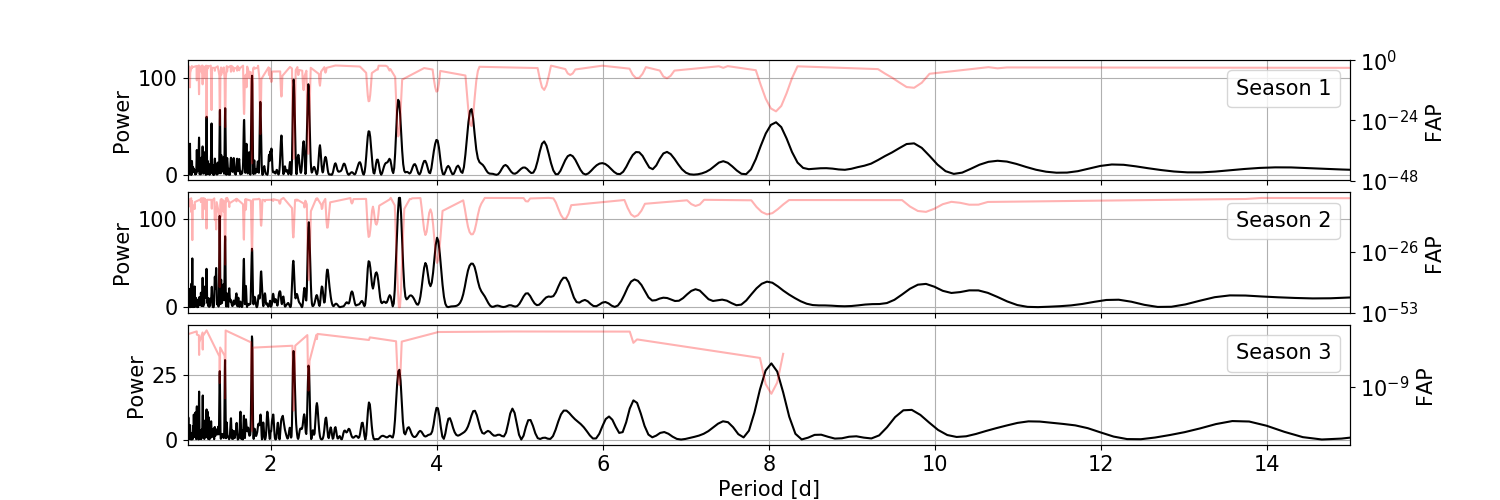}
    \includegraphics[width=\linewidth]{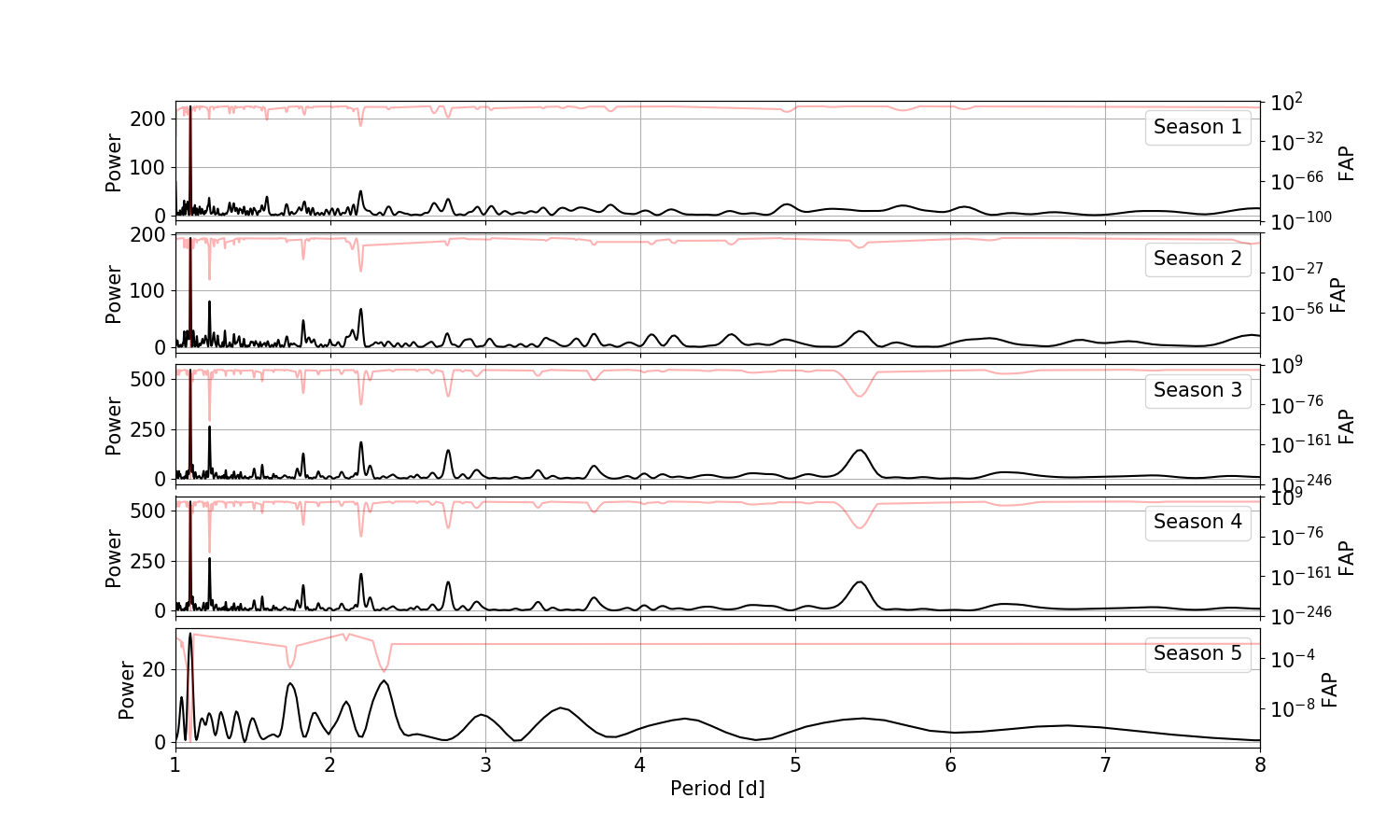}
    \includegraphics[width=\linewidth]{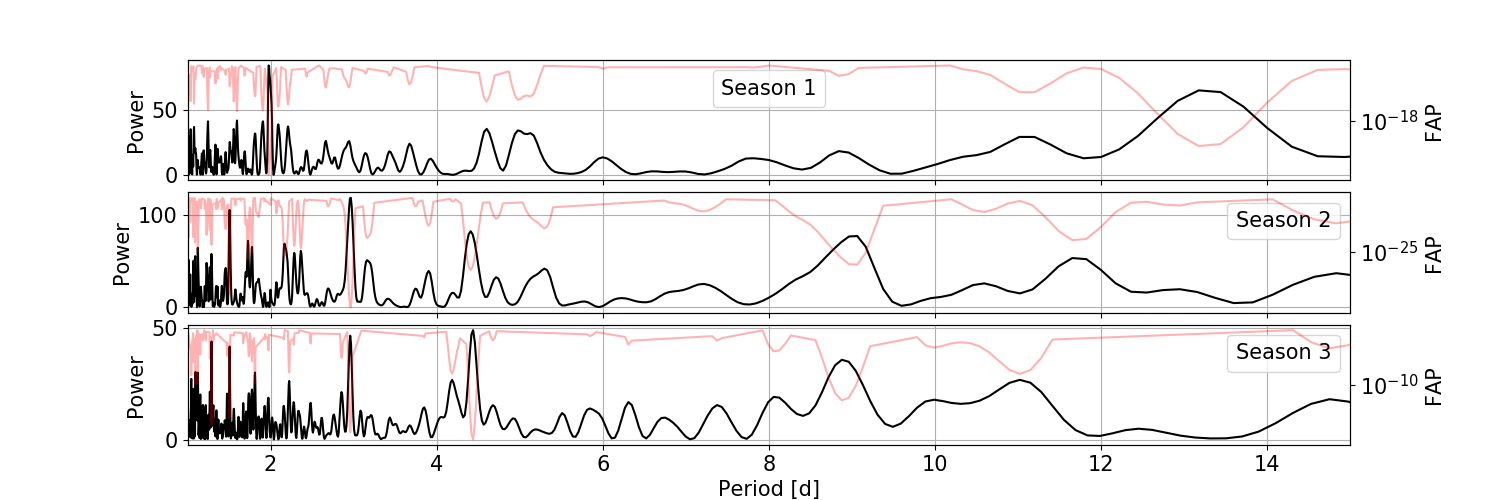}
    \caption{Lomb-scargle diagrams for each season of WASP photometry for J2349$-$32 (top), J2308$-$46 (middle) and J0218$-$31 (bottom).}
    \label{fig:my_label1}
\end{figure*}
\begin{figure*}
    \centering
     \includegraphics[width=\linewidth]{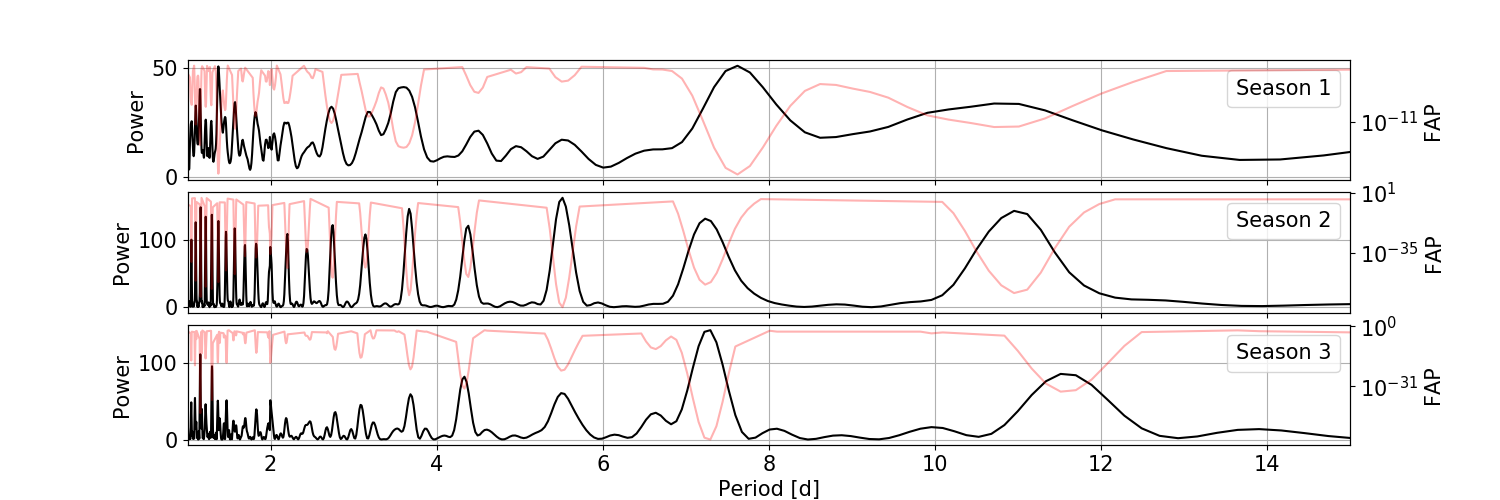}
    \includegraphics[width=\linewidth]{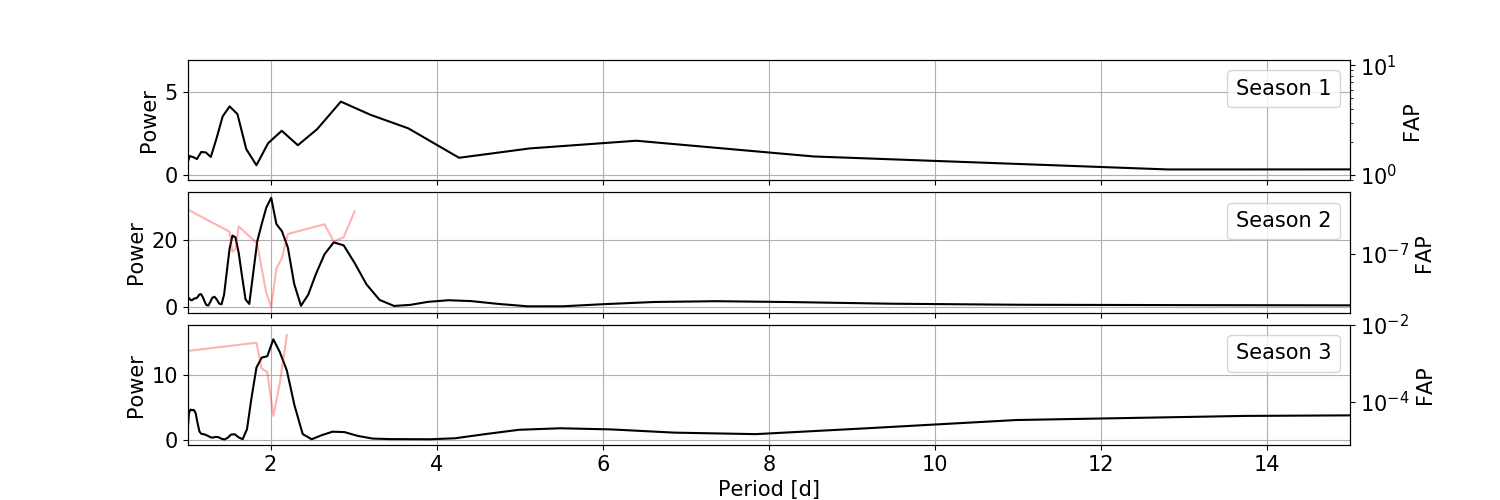}
    \caption{Lomb-scargle diagrams for each season of WASP photometry for J1847$+$39 (top) and J1436$-$13 (bottom).}
    \label{fig:my_label2}
\end{figure*}

\section{Spectroscopic observations}

\clearpage
\onecolumn
\begin{longtable}{lllrrr}
\caption{\label{appendix:spec_observations} Dates and times of spectroscopic observations. } 

\\ \hline
\hline 
Instrument   & Date [yyy-mm-dd] & Time [UT] \\
\hline

$J2349-32$   \\
CORALE &  2008-08-12  &  00:53:02 \\
CORALE &  2008-08-26  &  00:26:18 \\
CORALE &  2008-08-28  &  01:10:33 \\
CORALE &  2010-06-21  &  07:39:18 \\
CORALE &  2010-06-25  &  07:31:30 \\
CORALE &  2010-07-05  &  09:10:41 \\
CORALE &  2010-07-14  &  09:52:17 \\
CORALE &  2010-07-15  &  05:19:10 \\
CORALE &  2010-07-23  &  10:00:13 \\
CORALE &  2010-07-24  &  03:41:42 \\
CORALE &  2010-07-27  &  06:58:08 \\
CORALE &  2011-07-14  &  10:28:18 \\
CORALE &  2011-12-22  &  01:47:09 \\
CORALE &  2011-12-23  &  01:48:13 \\
CORALE &  2012-05-20  &  10:20:22 \\
CORALE &  2012-07-03  &  07:57:55 \\
CORALE &  2012-08-11  &  08:40:42 \\
CORALE &  2011-12-20  &  01:36:45 \\ \\

J2308-46 \\

CORALE &  2008-08-26  &  03:42:09 \\
CORALE &  2008-08-30  &  02:42:55 \\
CORALE &  2010-06-27  &  08:34:55 \\
CORALE &  2010-06-28  &  10:21:55 \\
CORALE &  2010-07-05  &  08:46:06 \\
CORALE &  2010-07-24  &  06:58:15 \\
CORALE &  2010-07-25  &  07:21:59 \\
CORALE &  2010-08-01  &  04:11:06 \\
CORALE &  2011-05-29  &  10:09:49 \\
CORALE &  2011-07-14  &  10:14:04 \\
CORALE &  2011-08-21  &  02:32:50 \\
CORALE &  2011-09-01  &  02:06:05 \\
CORALE &  2011-09-05  &  05:06:04 \\
CORALE &  2011-10-25  &  01:02:57 \\
CORALE &  2011-12-22  &  01:05:58 \\
CORALE &  2011-12-23  &  01:06:27 \\
CORALE &  2012-08-08  &  09:06:51 \\
CORALE &  2012-08-10  &  08:19:26 \\
CORALE &  2012-08-12  &  10:03:12 \\ \\

J0218-31 \\
CORALE &  2010-07-03  &  10:31:10 \\
CORALE &  2010-07-05  &  10:30:27 \\
CORALE &  2010-07-12  &  10:32:52 \\
CORALE &  2010-07-13  &  09:18:12 \\
CORALE &  2010-07-15  &  10:29:11 \\
CORALE &  2010-09-04  &  09:15:04 \\
CORALE &  2010-09-05  &  03:17:21 \\
CORALE &  2010-09-05  &  03:31:40 \\
CORALE &  2010-09-05  &  03:45:10 \\
CORALE &  2010-09-05  &  03:58:43 \\
CORALE &  2010-09-05  &  04:12:43 \\
CORALE &  2010-09-05  &  04:26:14 \\
CORALE &  2010-09-05  &  04:39:46 \\
CORALE &  2010-09-05  &  05:07:09 \\
CORALE &  2010-09-05  &  05:20:40 \\
CORALE &  2010-09-05  &  05:34:10 \\
CORALE &  2010-09-05  &  05:47:41 \\
CORALE &  2010-09-05  &  06:01:23 \\
CORALE &  2010-09-05  &  06:14:54 \\
CORALE &  2010-09-05  &  06:28:24 \\
CORALE &  2010-09-05  &  06:41:55 \\
CORALE &  2010-09-05  &  06:55:26 \\
CORALE &  2010-09-05  &  07:08:57 \\
CORALE &  2010-09-05  &  07:27:33 \\
CORALE &  2010-09-06  &  09:23:03 \\
CORALE &  2010-10-13  &  04:46:32 \\
CORALE &  2010-07-01  &  10:13:39 \\
CORALE &  2010-09-05  &  04:53:17 \\
CORALE &  2010-10-17  &  06:55:47 \\
CORALE &  2011-12-19  &  05:01:20 \\
CORALE &  2010-10-19  &  08:17:28 \\
CORALE &  2010-10-21  &  03:50:25 \\
CORALE &  2010-11-05  &  23:57:14 \\
CORALE &  2010-11-06  &  02:41:14 \\
CORALE &  2010-11-06  &  04:46:56 \\
CORALE &  2010-11-06  &  05:25:11 \\
CORALE &  2010-11-06  &  05:56:01 \\
CORALE &  2010-11-06  &  07:27:24 \\
CORALE &  2010-11-07  &  02:15:52 \\
CORALE &  2011-01-02  &  03:26:27 \\
CORALE &  2011-07-23  &  09:54:13 \\
CORALE &  2011-10-24  &  02:45:32 \\
CORALE &  2011-11-02  &  03:14:37 \\
CORALE &  2011-12-23  &  03:20:34 \\
CORALE &  2012-02-04  &  01:10:24 \\
CORALE &  2012-02-07  &  01:04:01 \\
CORALE &  2012-02-16  &  01:07:59 \\
CORALE &  2012-02-17  &  01:08:43 \\
CORALE &  2012-02-21  &  01:01:28 \\
CORALE &  2012-07-04  &  10:28:25 \\
CORALE &  2012-07-05  &  08:45:39 \\
CORALE &  2008-08-04  &  03:11:23 \\
CORALE &  2008-08-26  &  05:34:44 \\ \\

J1847+39 \\

INT &  2008-06-20  &  01:32:46 \\
INT &  2008-06-21  &  02:42:10 \\
INT &  2008-06-22  &  02:07:53 \\
INT &  2008-06-22  &  23:32:54 \\
INT &  2008-06-23  &  21:23:26 \\
INT &  2008-04-23  &  05:10:30 \\
INT &  2008-04-24  &  04:15:22 \\
INT &  2008-06-18  &  02:12:13 \\
INT &  2008-06-19  &  02:50:49 \\
INT &  2008-06-20  &  00:22:26 \\ \\

J1436-13  \\

CORALE &  2010-04-19  &  07:23:41 \\
CORALE &  2010-04-21  &  07:30:40 \\
CORALE &  2010-04-24  &  02:04:05 \\
CORALE &  2010-05-01  &  07:14:49 \\
CORALE &  2010-05-08  &  05:25:17 \\
CORALE &  2010-05-10  &  04:10:49 \\
CORALE &  2010-05-11  &  05:05:47 \\
CORALE &  2010-06-25  &  02:12:29 \\
CORALE &  2010-06-26  &  23:28:41 \\
CORALE &  2011-03-15  &  09:43:54 \\
CORALE &  2011-05-04  &  05:28:48 \\
CORALE &  2011-05-13  &  03:37:21 \\
CORALE &  2011-06-14  &  03:03:43 \\
\hline
\end{longtable}
\clearpage
\twocolumn
\end{appendix}

\end{document}